\documentclass{article}

\usepackage{PRIMEarxiv}
\usepackage[utf8]{inputenc}
\usepackage{amsthm}
\usepackage{multirow}
\usepackage{listings}
\usepackage{graphicx}
\usepackage{booktabs}
\usepackage[table]{xcolor}
\usepackage{algorithm}
\usepackage{algpseudocode}
\usepackage{subfigure}
\usepackage{mathtools}
\usepackage{amssymb}
\usepackage{enumitem}
\usepackage{lineno}

\newtheorem{definition}{Definition}
\usepackage[utf8]{inputenc} 
\usepackage[T1]{fontenc}    
\usepackage{hyperref}       
\usepackage{url}            
\usepackage{booktabs}       
\usepackage{amsfonts}       
\usepackage{nicefrac}       
\usepackage{microtype}      
\usepackage{lipsum}
\usepackage{fancyhdr}       
\usepackage{graphicx}       
\graphicspath{{media/}}     

\title{A Unified Active Learning Framework for Annotating Graph Data with Application to Software Source Code Performance Prediction 

}

\author{
  Peter Samoaa \\
  Chalmers University of Technology \\
  Data Science and AI \\
  \texttt{samoaa@chalmers.se} \\
   \And
   Linus Aronsson \\
  Chalmers University of Technology \\
  Data Science and AI \\
  \texttt{linaro@chalmers.se}  \\
  \And
  Antonio Longa \\
  Fondazione Bruno Kessler \\
  University of Trento, Italy\\
  alonga@fbk.eu\\
  \And
  Philipp Leitner \\
  Chalmers University of Technology \\
  Interaction Design and Software Engineering \\
  philipp.leitner@chalmers.se \\
  \And
  Morteza Haghir Chehreghani \\
  Chalmers University of Technology \\
  Data Science and AI \\
  \texttt{morteza.chehreghani@chalmers.se} \\
}

\begin{document}
\maketitle

\begin{abstract}
Most machine learning and data analytics applications, including performance engineering in software systems, require a large number of annotations and labelled data, which might not be available in advance. Acquiring annotations often requires significant time, effort, and computational resources, making it challenging. We develop a unified active learning framework specializing in software performance prediction to address this task. We begin by parsing the source code to an Abstract Syntax Tree (AST) and augmenting it with data and control flow edges. Then, we convert the tree representation of the source code to a Flow Augmented-AST graph (FA-AST) representation. Based on the graph representation, we construct various graph embeddings (unsupervised and supervised) into a latent space. Given such an embedding, the framework becomes task agnostic since active learning can be performed using any regression method and query strategy suited for regression. Within this framework, we investigate the impact of using different levels of information for active and passive learning, e.g., partially available labels and unlabeled test data. Our approach aims to improve the investment in AI models for different software performance predictions (execution time) based on the structure of the source code. Our real-world experiments reveal that respectable performance can be achieved by querying labels for only a small subset of all the data.
\end{abstract}


\section{Introduction}

In this paper, we investigate the use of \emph{active learning} to annotate the graph representation of program source code, with the goal of predicting execution times prior to execution. Active learning is a practical learning approach that effectively finds the most informative data points for labelling instead of annotating all of them \cite{settles.tr09}. Applying Artificial Intelligence (AI) to software engineering is limited due to a lack of annotated data~\cite{samoaaSLR2022}. Millions of open source projects are available on public host platforms such as Github (each with dozens, hundreds, or thousands of source code files to learn from), but without proper labelling for specific software development tasks, such as performance prediction. Performance prediction is the process of predicting non-functional properties such as the execution time of the source code before it runs. The benefit of performance prediction is that it can give developers an early indication about performance issues without the need for expensive testing.
However, collecting performance data is costly, as it requires executing many applications using many different workloads (input or runtime configuration), ideally in a controlled test environment. This consumes a lot of time, effort and computational resources.


To deal with this challenge, we develop a generic active learning framework suitable for graph data and continuous target variables (i.e., regression tasks). Active learning has been successfully applied to several tasks such as image processing \cite{sener2018active,LiO21,BosserSC21,CasanovaPRP20}, recommender systems \cite{Rubens2015}, driver behaviour identification \cite{ComuniMAC22}, sound event detection \cite{TASLP.2020.3029652}, classification of driving time series \cite{JARL2022104972}, reaction prediction in drug discovery \cite{minf.202200043}, logged data analysis \cite{YanCJ18}, medical analysis \cite{NIPS2017_8ca8da41,LiO21}, text processing \cite{vu-etal-2019-learning}, and person re-identification \cite{LiuWGTL19}. Several studies have demonstrated the effectiveness of active learning on graphs where the data items correspond to the nodes of a \emph{single} graph, e.g., \cite{cai2017,Wu2019,hu2020graph,Zhang2022}.
However, to our knowledge, there is no prior work on active learning on graph representation for the source code. Moreover, There is still no maturity in the investment of active learning on graph level where each data item has its own graph, i.e., $N$ different graphs represent the $N$ data items in the dataset.


However, dealing with graph data is crucial in various real-world applications. In performance prediction, each source code file is represented as an abstract syntax tree (AST). Following our previous work~ \cite{SamoaaTEPGNN}, one can achieve a richer representation by augmenting the AST by adding multiple edge types that describe the semantics of the sequence of execution or how the code is executed. Thus, the tree structure of the represented source code is converted to a graph called Flow Augmented-AST (FA-AST).


In this paper, for the first time, we develop an active learning framework applicable to graph representation for the source code where the learning is done on a graph level. Our framework can be, in principle, applied to any kind of graph data. Within this framework, we first construct an embedding that maps each graph into a latent space representation that captures the complex properties of the graph. We consider a wide range of graph embedding techniques, both unsupervised and supervised. Given such an embedding, we are no longer limited to (regression) models that can directly handle graph data. Instead, any regression model can be utilized. In turn, this gives rise to the possibility of using any existing active learning query strategy suited for regression. In this paper, we investigate Gaussian Process Regressors (GPR)~\cite{GPR} as they give rise to  natural acquisition functions for active learning in the regression setting. We consider four different acquisition functions in the active learning experiments: uniform random, Coreset~\cite{coreset2021}, variance~\cite{varaince_2017} and query-by-committee(QBC)~\cite{QBC2019} (see Section \ref{section:active-learning-strategy} for details).



In addition, we investigate the effectiveness of using different information when constructing different graph embeddings for active learning. Such information could, for example, correspond to the features of test data (not the labels) and/or the partially known labels (acquired by active learning) for training  data. Overall, our study provides valuable insights into using additional information for active (and passive) learning with graph embeddings and sheds light on the potential benefits of leveraging different types of information in this context.


Finally, we investigate the framework on the real-world application of performance prediction based on source code, where the availability of annotated data has traditionally been limited. To our knowledge, no mature active learning method exists for source code representation for performance prediction. Thus, our study fills an important gap in the literature and offers a practical solution to this challenging problem.
Our experiments show that active learning can help achieve good performance prediction with only a subset of all available labels. The results show that uniform random in most cases is the worst. Whereas  QBC, variance-based, and Coreset alternate in order of preference according to the dataset, setting of the experiment, and quality of embeddings and used embedding algorithms. In addition, the results indicate that Graph2Vec~\cite{graph2vec} (which is entirely unsupervised) can construct powerful graph embeddings, sometimes comparable to embeddings constructed by a supervised Graph Neural Network(GNN)~\cite{SamoaaTEPGNN}.


Thus, in summary, our contributions consist of i) a unified framework for active learning on graph data, ii) investigation of the impact of various additional information on active learning, and iii) application of the framework to the novel real-world task of software performance prediction. This research opens the door for more investment in AI for software performance engineering research, providing a practical and efficient approach for annotation and labelling source code data for performance prediction.

Our work provides a novel open-source framework enabling researchers to investigate various active learning methods for graph data. The code is publicly available at~\cite{peter_samoaa_2023_7792485}.

\section{Background} \label{section:background-appendix}
\subsection{Source Code Representation}
\begin{figure*}[htb!] 
\centering 
\subfigure[Abstract syntax tree (AST)  for the code snippet in Listing 1~\cite{Yamaguchi2014}.]{\label{fig:AST}\includegraphics[width=0.6\linewidth]{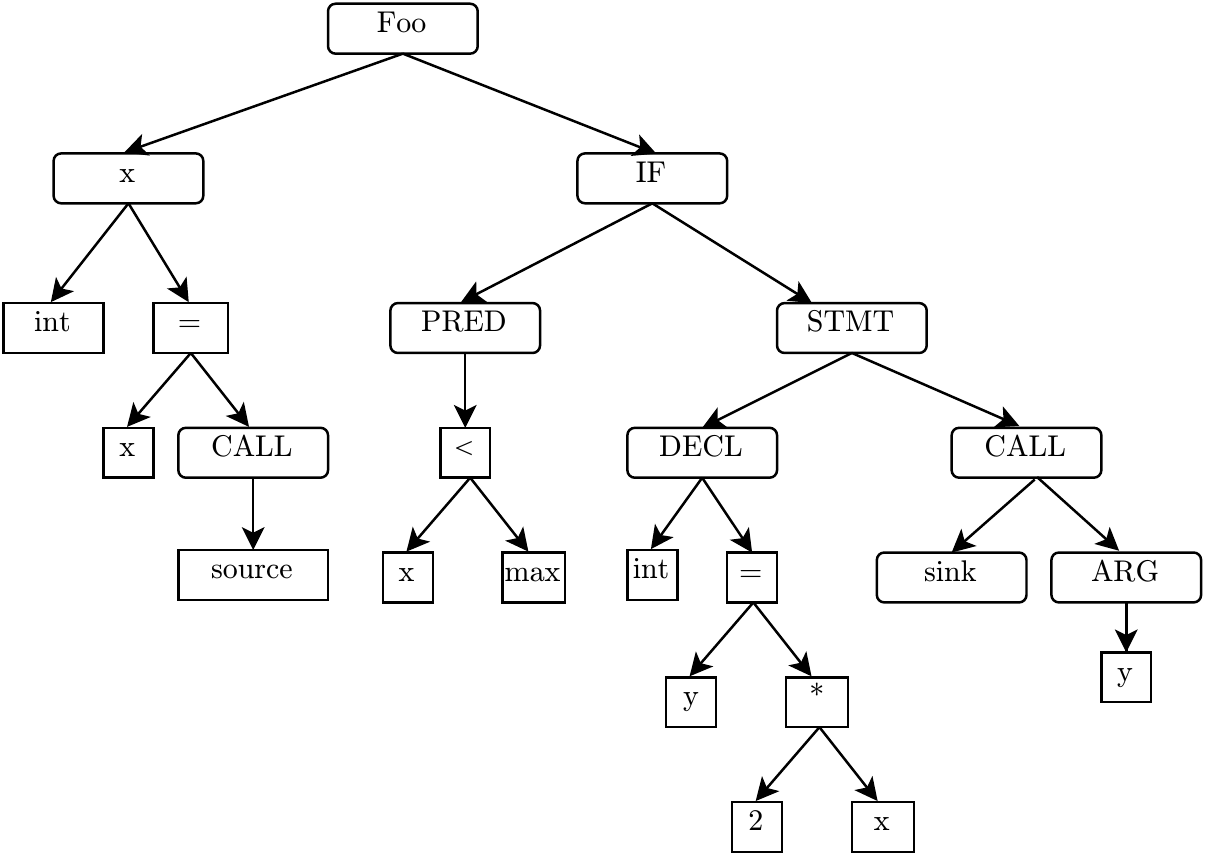}}
\hfill
\subfigure[Control Flow Graph (CFG) for the code snippet in Listing 1~\cite{Yamaguchi2014}.]{\label{fig:CFG}\includegraphics[width=0.35\linewidth]{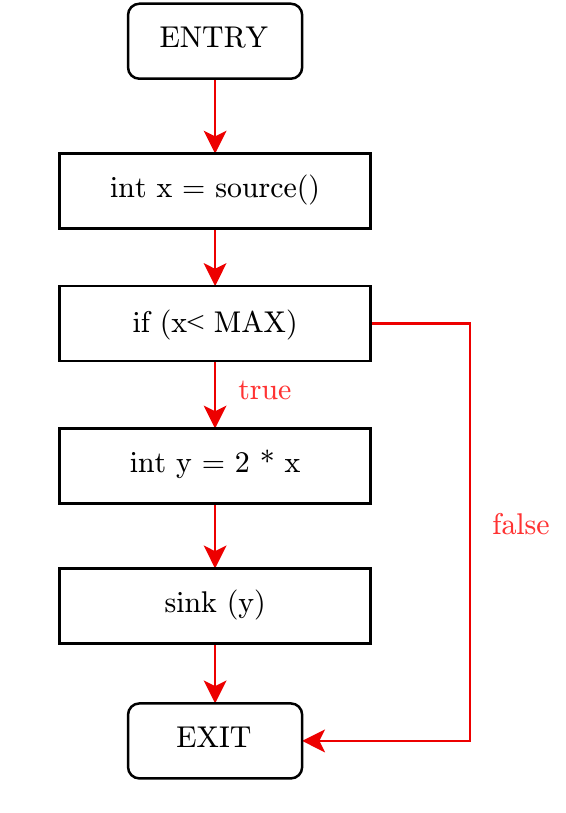}}

\caption{Example of Tree and Graph Representation for the code snippet in Listing 1 (from Yamaguchi et al.~\cite{Yamaguchi2014}).}
\end{figure*}

This study aims to increase the amount of annotated data cost-efficiently for performance prediction based on the structure of the source code, making it necessary to provide a comprehensive background on the intermediate representation of code. As outlined in our earlier mapping study~\cite{samoaaSLR2022}, program source code can be represented using tree-based, graph-based, or token-based approaches. Among these, the tree-based approach, specifically the Abstract Syntax Tree (AST) as shown in Figure~\ref{fig:AST}, provides valuable information about the code structure through its syntactical and lexical information. Unlike other graph-based approaches, such as data flow graphs (DFG), the AST representation can be extracted through source code parsing alone, without the need for executing the program. In this study, we rely on the AST representation due to its ease of extraction through source code parsing and the abundance of structural information it provides. Additionally, it should be noted that graph neural networks require many edges to extract meaningful information, and the AST representation provides many nodes, making it well-suited for this task. 

 \begin{lstlisting}[float=h, language=C, caption=Simple example of C source code (from Yamaguchi et al.~\cite{Yamaguchi2014})., label=c:example]
 void foo() {
   int x = source();
   if( x < MAX ) {
     int y = 2*x;
     sink(y);
   }
 }
 \end{lstlisting}

%

\subsection{Graphs}
A graph is a mathematical structure used to model relational data. Graphs have been used to model several domains, i.e.,  biology \cite{aittokallio2006graph,huber2007graphs}, face-to-face human interactions \cite{longa2022efficient,longa2022neighbourhood}, social network analysis~\cite{scott2011social,nguyen2022emotion,lachi2023impact}, mobility networks\cite{cho2011friendship,mauro2022generating}, digital contact tracing \cite{cencetti2021digital}, or software representation \cite{SamoaaTEPGNN,suneja2020learning}. Below we are giving some definitions used later in the paper.

\begin{definition}[\textit{Graph}]
  A graph $G$, in general, can be defined as a pair $(V,E)$, where $V$ is a set of vertices or nodes, and $E$ is a set of edges between the nodes, i.e., $E \subseteq \{(u,v) | u,v \in V\}$. \\
  A graph can be represented with a squared adjacency matrix $A$ of size $|V|\times |V|$ in which the element $a_{i,j}$ is one if the graph has an edge from $i$ to $j$. The graph is undirected if it does not contain self-loops and the associated adjacency matrix is symmetric, directed otherwise. At each node, (or edge) of the graph can be associated features.
\end{definition}

\begin{definition}[\textit{Path}]
    A path $P = \{v_1,\dots,v_k\}$ is an ordered sequence of connected nodes on the graph. The length of a path is the number of nodes in the path. Given two nodes $u,v$ the shortest path is the path with the minimal length connecting node $u$ and node $v$.
\end{definition}

\begin{definition}[\textit{Node neighborhood}]
  Given a  graph $G = (V,E)$, the {\em neighbors} of a node
  $v \in V$ are the set of nodes adjacent to $u$, i.e.,
  $\text{Neighbors}(v) = \{u \in V | (u,v) \in E\}$. The node {\em neighbourhood} is
  the subgraph of $G$ containing $v$ and its neighbours as nodes and
  all edges connecting them as edges.
\end{definition}

\begin{definition}[\textit{Degree}]
    The degree of a node $v$ is the number of neighbours of the node.
\end{definition}

\begin{definition}[\textit{Density}]
    The density of a directed graph is defined as:
    $$\text{Density} = \dfrac{|E|}{|V| (|V| - 1)}$$
\end{definition}

\begin{definition}[\textit{Open and closed triads}]
  Given a  graph $G = (V, E)$, a triad is a subset of $V$ with three connected nodes. If the triad has three edges, (i.e. it is a triangle) then the triad is closed, open otherwise.
\end{definition}

\section{Related Work}

Active Learning (AL) has been extensively explored across fields like text~\cite{shen2018} and image~\cite{Gal2017} data to improve data annotation procedures in these domains, leading to more practical AI applications. For graph data, AL has been effectively employed in densely connected graphs~\cite{abel2019,Li2022}. However, the effectiveness of AL on sparse graphs remains an open research question.

Cai et al.~\cite{cai2017} proposed AGE, an active graph embedding framework that operates on the node level and uses uncertainty and representativeness as query strategies. Wu et al.~\cite{Wu2019} introduced a generic active learning framework using distance-based clustering. Both studies utilized GCN for node representation learning.

Hu et al.~\cite{hu2020graph}, and Zhang et al.~\cite{Zhang2022} applied active learning to graph data by employing reinforcement learning. Hu et al. proposed a Graph Policy Network for transferable active learning on graphs called GPA, which formalizes active learning on graphs as a Markov decision process (MDP) and learns the optimal query strategy with reinforcement learning. In this approach, the action selects a node for annotation at each query step, while the state is defined based on the current graph status. The reward is the performance gain of the GNN trained with the selected nodes. In contrast, Zhang et al. formulated BIGENE, a batch active learning method, as a cooperative multi-agent reinforcement learning problem.

Gao et al.~\cite{Gao2018}, and Chen et al.~\cite{Chen2019} investigated multi-arm bandits in an active learning setting. Gao et al. proposed ANRMAB for learning discriminative network representations and used Information Entropy, Node Centrality, and Information Density as query strategies for node-level labelling. Meanwhile, Chen et al. proposed ActiveHNE for heterogeneous network embedding and combined Network Centrality, Convolutional Information Entropy, and Convolutional Information Density as a selection strategy based on uncertainty and representativeness.

Three common aspects among all the aforementioned studies are: a) they did not use real-world datasets but rather benchmark datasets such as Citeseer, Cora, and Pubmed to validate their methods, b) they employed semi-supervised learning, and c) they focused on the node level. In contrast, our approach utilizes real-world datasets, operates on both node and graph levels, and incorporates supervised, and unsupervised learning.

Finally, the generic active learning framework introduced by Jarl et al.~\cite{JARL2022104972} will also be followed in this paper. However, the previous work uses temporal data, and in our study, we deal with graph data. Additionally, we perform experimental studies that more systematically investigate the performance of this framework with different levels of information.

\section{Learning Framework}

\begin{figure}
    \centering
    \includegraphics[scale = 0.45]{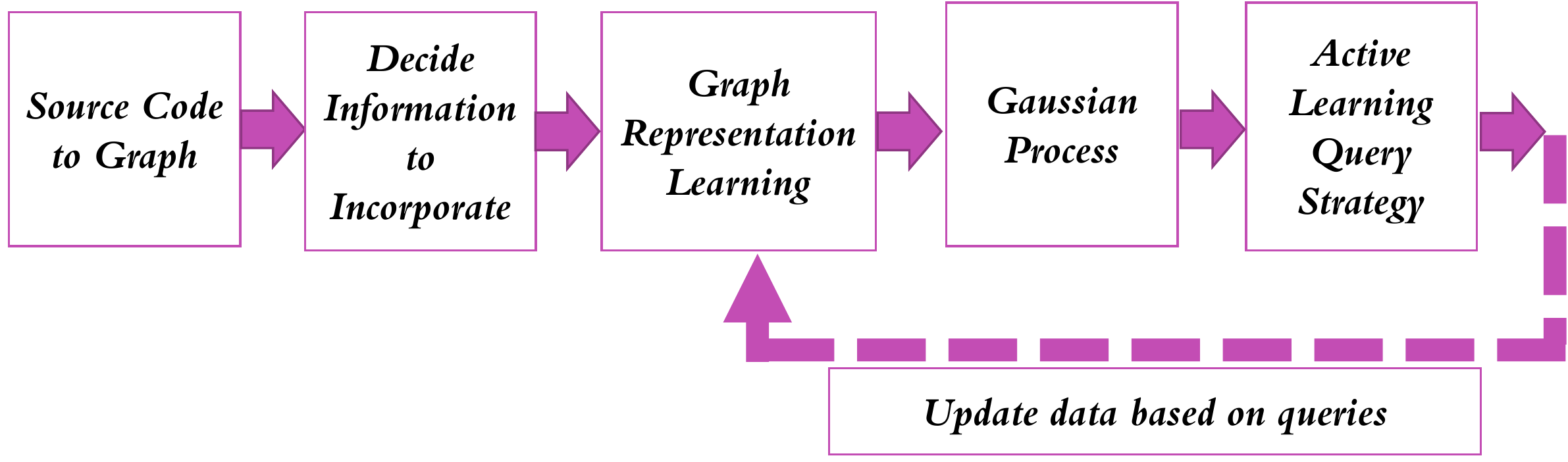}
    \caption{Representation learning and Active Learning Strategies}
    \label{fig:ALframework}
\end{figure}

In this section, we provide a detailed overview of our active learning framework. 
Figure~\ref{fig:ALframework} presents our framework for active learning. Section \ref{section:al-procedure} begins by explaining the general setup for active and passive learning given a graph dataset. The remaining sections will then explain each of the components visualized in Figure~\ref{fig:ALframework}.

\subsection{Active and Passive Learning Procedure} \label{section:al-procedure}

We are given a dataset $\mathcal{D}$ of $N$ source code files (represented as graphs, see next section). For active learning, we then split this dataset into three parts, the initially labelled dataset $\mathcal{L}_0$, the initially unlabeled dataset $\mathcal{U}_0$ and a test set $\mathcal{T}$. The purpose of the test set is to be able to evaluate the active learning procedure. Active learning can be seen as an iterative procedure where in each iteration $i$, one begins by training some regressor $\mathcal{R}_i$ based on the currently available information, i.e., $\mathcal{L}_i$, $\mathcal{U}_i$ and possibly $\mathcal{T}$.\footnote{Note that here we assume for the test data only the data features might be available to be utilized, not the labels. Assume, for example, a photographer has taken two sets of photos from the same objects. For the first set (i.e., the training dataset) she has the image labels, but for the second set (i.e., the test dataset) only the images (without labels) are available. When training a classifier, she may then use the test images as well, in addition to labeled training dataset.} Then, the current regressor $\mathcal{R}_i$ is evaluated using the test set $\mathcal{T}$. Then, a query strategy is used to select the most informative batch $\mathcal{B} \subseteq \mathcal{U}_i$ of data items from $\mathcal{U}_i$ based on information in the following components: $\mathcal{R}_i$, $\mathcal{L}_i$, $\mathcal{U}_i$ and $\mathcal{T}$. Finally, the datasets are updated by setting $\mathcal{L}_{i+1} \coloneqq \mathcal{L}_i \cup \mathcal{B}$ and $\mathcal{U}_{i+1} \coloneqq \mathcal{U}_i \setminus \mathcal{B}$. This is repeated until a stopping criterion is met (e.g., if the labelling budget has been reached). In addition to active learning, we conduct experiments in the passive setting, which corresponds to setting $\mathcal{L} = \mathcal{L}_0$ and $\mathcal{U}_0 = \emptyset$. Then, one trains a regressor $\mathcal{R}$ on $\mathcal{L}$ and makes predictions on $\mathcal{T}$ (i.e., the traditional supervised machine learning).


\subsection{Transforming Source Code to Graphs}

\begin{figure}
    \centering
    \includegraphics[scale = 0.5]{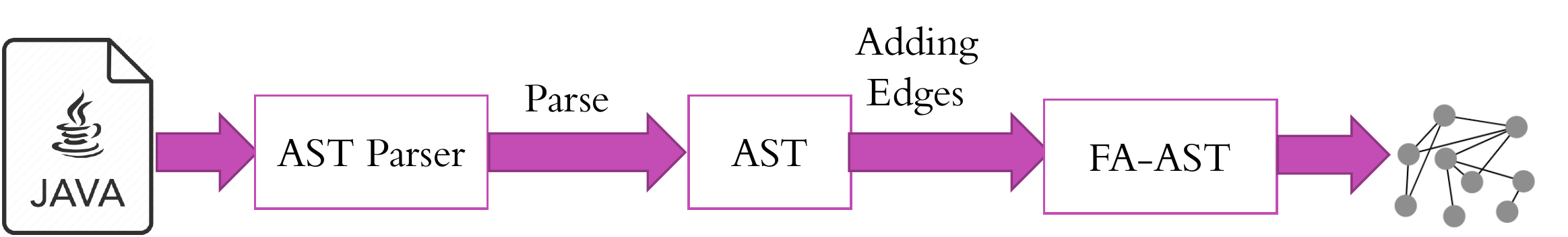}
    \caption{Source Code to Graph Process}
    \label{fig:SC2G}
\end{figure}

This section explains how to build the graphs from the source codes. 
As shown in Figure~\ref{fig:SC2G}, we investigate Java source code files. We represent the source code as an AST intermediate representation. To compress both semantic and syntactical information, we augment the AST by adding edges that preserve both data and control the flow of the graphs. Hence, we arrive at a flow-augmented AST (FA-AST) graph, a concept that we introduced in our earlier work~\cite{SamoaaTEPGNN}.

Our motivation for augmenting the AST comes from recent studies \cite{samoaaSLR2022}, emphasising the importance of rich code representation when using deep learning in software engineering. Hence, and
given the complexity of predicting performance, prediction based on the syntactical information extracted from ASTs alone is not sufficient to achieve high-quality predictions. The AST's basic structural information is enriched with semantic information representing data and control flow. Consequently, the tree structure of the AST is generalized to a (substantially richer) graph, encoding more information than the code structure alone. 

\subsubsection{Motivation Example}
To understand how the graphs are built, we will present an example for a Java code file and then explain in detail how the FA-AST is built (see Listing~\ref{java:example}). 

\begin{lstlisting}[float=h!, language=Java, caption=A Simple JUnit 5 Test Case, label=java:example, basicstyle=\scriptsize]
package org.myorg.weather.tests;

import static
    org.junit.jupiter.api.Assertions.assertEquals;
import org.myorg.weather.WeatherAPI;
import org.myorg.weather.Flags;

public class WeatherAPITest {
    
    WeatherAPI api = new WeatherAPI();
    
    @Test
    public void testTemperatureOutput() {
        double currentTemp = api.currentTemp();
        Flags f = api.getFreezeFlag();
        if(currentTemp <= 3.0d)
            assertEquals(Flags.FREEZE, f);
        else
            assertEquals(Flags.THAW, f);
    }
}
\end{lstlisting}

\paragraph{\textbf{AST Parsing}}
In this  example, a single test case \texttt{testTemperature\-Output()} is presented that tests a feature of an (imaginary) API. As common for test cases, the example is short and structurally relatively simple. Much of the body of the test case consists of invocations to the system-under-test and calls of JUnit standard methods, such as \texttt{assertEquals}.

\begin{figure}[h!]
    \centering
    \includegraphics[width=\textwidth]{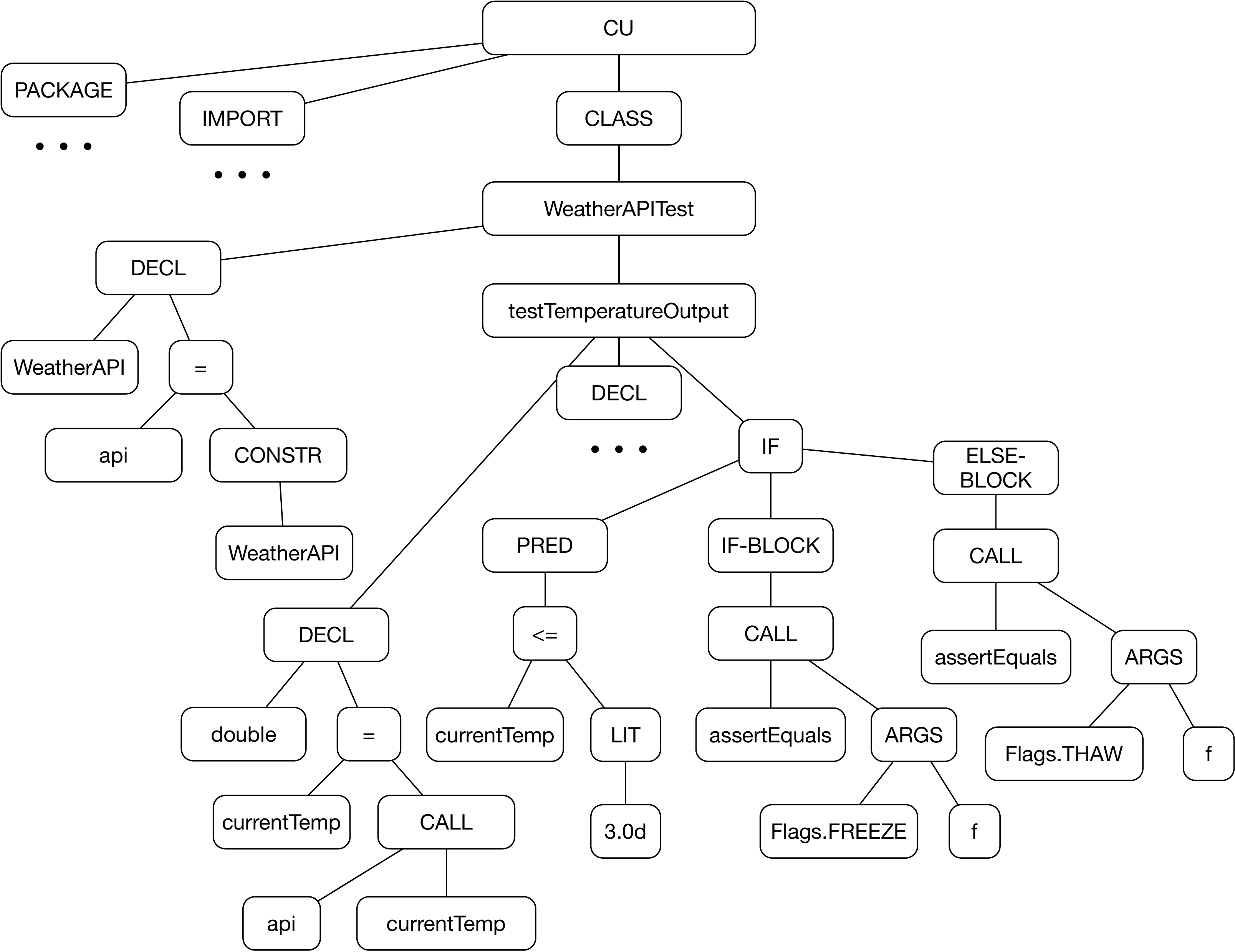}
    \caption{Simplified abstract syntax tree (AST) representing the illustrative example presented in Listing~\ref{java:example}. Package declarations, import statements, as well as the declaration in Line 15 are skipped for brevity.}
    \label{fig:AST_draft}
\end{figure}

A (slightly simplified) AST for this illustrative example is depicted in Figure~\ref{fig:AST_draft}. The produced AST does not contain purely syntactical elements, such as comments, brackets, or code location information. We make use of the pure Python Java parser javalang\footnote{https://pypi.org/project/javalang/} to parse each test file and use the node types, values, and production rules in javalang to describe our ASTs.

\vspace{1cm}

\paragraph{\textbf{Capturing Ordering and Data Flow}}

In the next step, we augment this AST with different types of additional edges representing data flow and node order in the AST. Specifically, we use the following additional flow augmentation edges, in addition to the \textbf{AST child} and \textbf{AST parent} edges that are produced readily by AST parsing:

\textbf{FA Next Token} (b):\\
This type of edge connects a terminal node (leaf) in the AST to the next terminal node. Terminal nodes are nodes without children. In Figure~\ref{fig:AST_draft}, an FA Next Token edge would be added, for example, between \texttt{WeatherAPI} and \texttt{api}.

\textbf{FA Next Sibling} (c):\\
This connects each node (both terminal and non-terminal) to its next sibling, and allows us to model the order of instructions in an otherwise unordered graph. In Figure~\ref{fig:AST_draft}, such an edge would be added, for example, connecting the first usage of \texttt{api} and with the \texttt{CONSTR} node (representing a Java constructor call).

\textbf{FA Next Use} (d):\\
This type of edge connects a node representing a variable to the place where this variable is next used. For example, the variable \texttt{api} is declared in Line 10 in Listing~\ref{java:example}, and then used next in Line 14. \begin{figure}
    \centering
    \includegraphics[width=\textwidth]{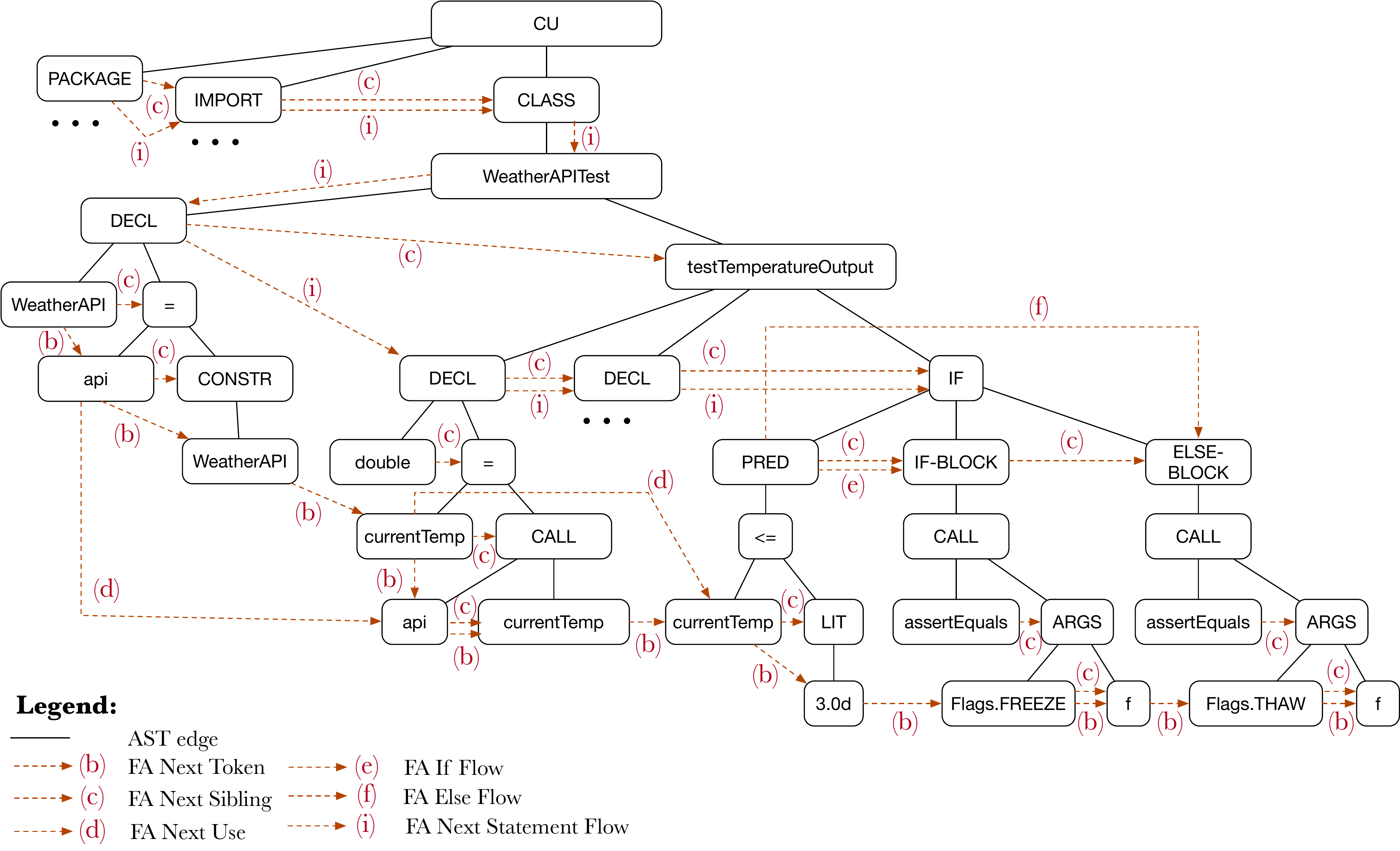}
    \caption{Flow-Augmented AST (FA-AST) for the example presented in Listing~\ref{java:example}. Solid lines represent AST parent and child edges, and dashed lines different types of flow augmentations.}
    \label{fig:FAAST}
\end{figure}

Figure \ref{fig:FAAST} shows an example augmenting the AST in Figure \ref{fig:AST} (and, consequently, the example test case in Listing~\ref{java:example}). Solid black lines indicate the AST parent and child relationships (for simplicity indicated through a single arrow, read from top to bottom). Red dashed arrows refer to the new edges added to represent the data and control flow in the FA-AST, with letter codes indicating the edge type. Terminal nodes are connected with FA Next Token edges (b), modelling the order of terminals in the test case. Similarly, the ordering of siblings is modelled using FA Next Sibling edges (c). Finally, data flow is modelled by connecting each variable to their next usage via FA Next Use edges (d). Edge types (e), (f), and (i) represent a control flow statement, which will be discussed in the following. Multiple edges of different types are possible between the same nodes. For example, the terminal nodes \texttt{Flags.FREEZE} and \texttt{f} are connected via both, an FA Next Token (b) and an FA Next Sibling (c) edge.

\vspace{-0.5cm}

\begin{figure}[h!]
    \centering
    \includegraphics[width=\columnwidth]{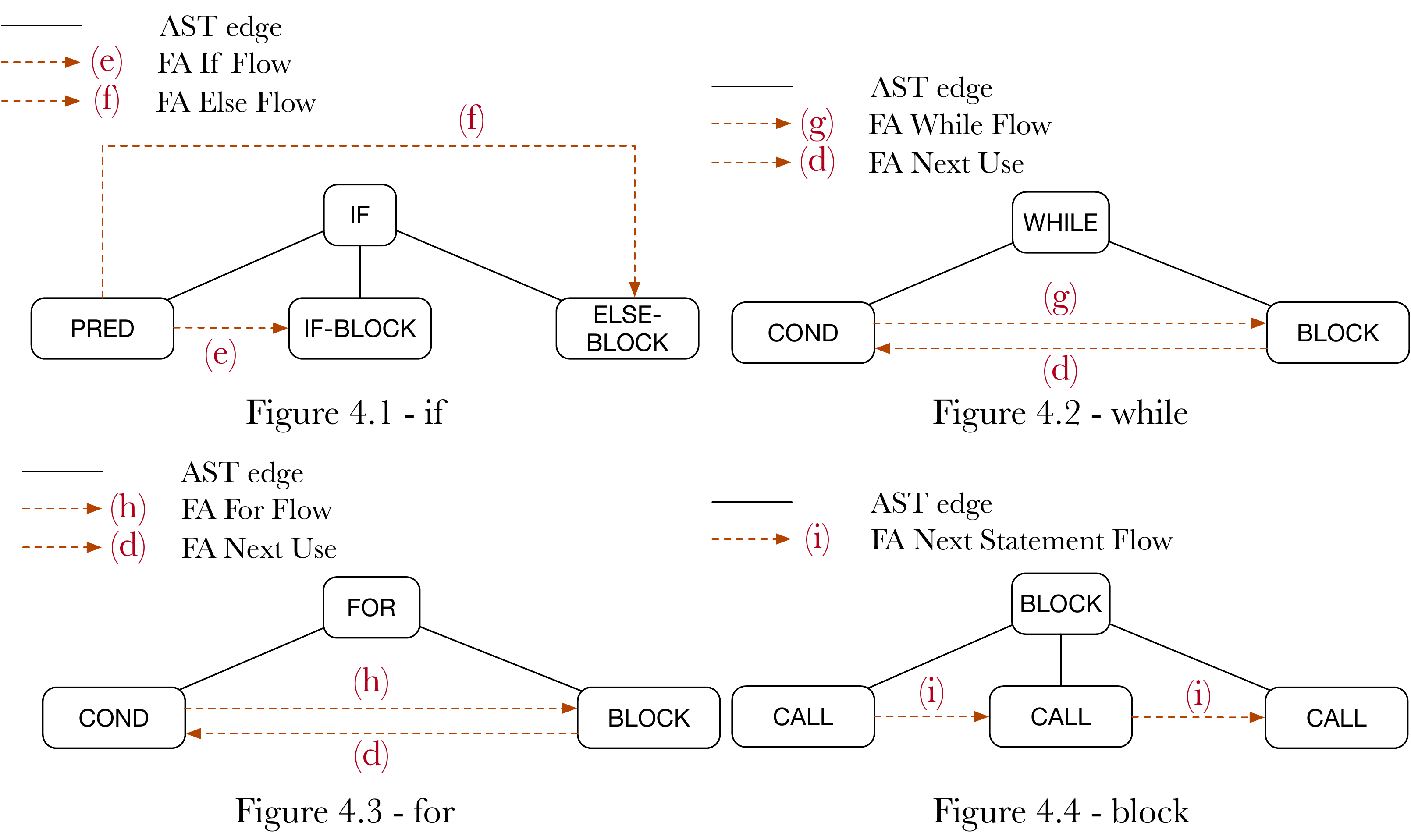}
    \caption{Additional flow augmentations for different control flow constructs}
    \label{fig:FA}
\end{figure}

\paragraph{\textbf{Capturing Control Flow}}
In a second augmentation step, we now add further edges representing the control flow in the test cases. We currently support \emph{if} statements, \emph{while} and \emph{for} loops, as well as \emph{sequential execution}. We currently do not support \emph{switch} statements or \emph{do-while} loops, as these are less common. Java source code containing these elements will still be parsed successfully, but these control flow constructs will not be captured by the FA-AST. Specifically, the following further edges are added (see also Figure~\ref{fig:FA}):

\textbf{FA If Flow} (e):\\
This type of edge connects the predicate (condition) of the if-statement with the code block that is executed if the condition evaluates to \texttt{true}. Every if-statement contains exactly one such edge by construction.

\textbf{FA Else Flow} (f):\\
Conversely, this edge type connects the predicate to the (optional) else code block.

\textbf{FA While Flow} (g):\\
A while loop essentially entails two elements - a condition and a code block that is executed as long as the condition remains \texttt{true}. We capture this through a FA While Flow (g) edge connecting the condition to the code block, and an FA Next Use (d) edge in the reverse direction. The latter is used to model the next usage of a loop counter. 

\textbf{FA For Flow} (h):\\
For loops are conceptually similar to while loops. We use FA For Flow (h) edges to connect the condition to the code block, and an FA Next Use (d) edge in the reverse direction. Similar to the modelling of while-loops, FA Next Use (d) relates to the usage (typically incrementing) of a loop counter.

\textbf{FA Next Statement Flow} (i):\\
In addition to the control flow constructs discussed so far, Java of course also supports the simple sequential execution of multiple statements in a sequence within a code block. FA Next Statement Flow edges (i) are used to represent this case. Different from the constructs discussed so far, a code block can contain an arbitrary number of children, and the FA Next Statement Flow edge is always used to connect each statement to the one directly following it.

Referring back to Figure \ref{fig:FAAST}, two types of control flow annotations are visible: the modelling of the if-statement in lines 16 to 19 of the test case on the right-hand side and various edges representing sequential executions (FA Next Statement flow (i) ). Further note how flow annotation adds a large number of edges to even a very small AST, transforming the syntax tree into a sparse graph. This rich additional information can be used in the next step by our GNN model to predict highly accurate test execution times.

\subsection{Depth of FA-AST Parsing}

One challenge with representing source code as graphs is that graphs tend to become very large. We address this challenge by limiting how deeply we parse the AST. We investigate two alternatives:

\begin{itemize}

\item \textbf{File-Level Parsing:} in the first alternative, we parse the AST only on the level of individual Java source files. References to Java constructs (e.g., classes, functions, etc.) not implemented in this file are turned into leaf nodes (and not resolved further). This leads to graphs of manageable size and has the added benefit of simplifying parsing, but evidently much expressive information is lost.

\item \textbf{System-Level Parsing:} in the second alternative, the parser has access to all source code files of the study subject system (e.g., all source code files of Hadoop when constructing FA-ASTs for Hadoop), and the all references to classes or functions that are implemented in the study subject are resolved fully. External dependencies or calls to the Java system library are not resolved, these remain represented as leaf nodes. This parsing strategy leads to substantially larger and more complex graphs, but has the benefit that more knowledge about the performance of methods of the study subject is represented in the graph.

\end{itemize}


\subsection{Graph Representation Learning}

The graph structure of the data items in $\mathcal{D}$ yields a restriction on the types of regression models that can be used, and thus the types of query strategies to use for active learning. Therefore, we investigate a number of unsupervised and supervised approaches to constructing embeddings that can be used to project the graph data into a latent space where any regression model (and thus query strategy) can be used. In this section, we outline each of the embeddings that we investigate in this work.

Since our focus is on directed graphs, we use embedding algorithms compatible with directed graphs where the adjacency matrix is not symmetric. For this purpose, we explore three main approaches: unsupervised embeddings (based on Graph Neural Networks (GNNs) and shallow embedding algorithms), supervised embeddings (based on GNNs) and manual embeddings (based on manually extracted graph features). Each of these categories are listed and explained below.


\subsubsection{Unsupervised embeddings.} 
\begin{figure}
    \centering
    \includegraphics[width=0.8\textwidth]{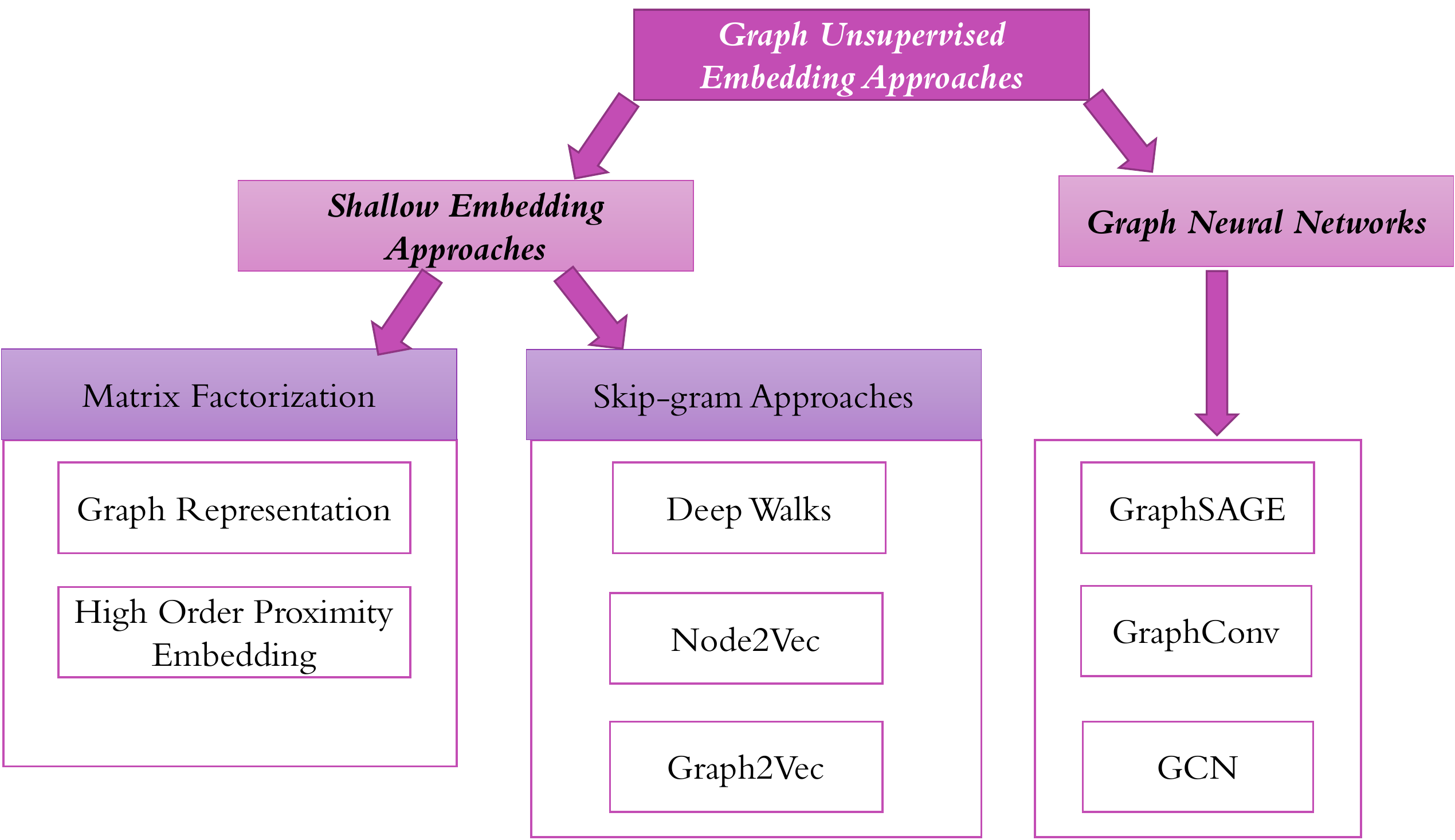}
    \caption{Hierarchical structure of the different unsupervised graph embedding algorithms used in this study.}
    \label{fig:unsupervised}
\end{figure}

Figure~\ref{fig:unsupervised} illustrates the hierarchy of unsupervised embedding algorithms used. The hierarchy is inspired by Chami et al.~\cite{Chami2022}. We have two main types of shallow embedding approaches: matrix factorization and skip-gram. In matrix factorization, we use the Graph Representation approach~\cite{GraRep} and Higher-Order Proximity Preserved Embedding (HOPE)~\cite{HOPE}, both of which are compatible with directed graphs. These algorithms operate at the node level, resulting in an embedding array for each graph rather than a vector. Therefore, we aggregate the embedding using mean and sum aggregation to represent the graph embeddings as vectors. For skip-gram related methods, we use DeepWalk~\cite{DW}, Node2Vec~\cite{node2vec}, both of which learn the embedding at the node level, and Graph2Vec, which creates Weisfeiler-Lehman tree features for nodes in graphs. A graph feature co-occurrence matrix is decomposed to generate graph representations using these features. Consequently, this is the only method for the shallow embedding category that returns a vector representing the embedding for the entire graph. According to Chami et al.~\cite{Chami2022}, shallow embedding methods are applied to a finite set of input graphs and cannot be applied to instances different from those used to train the model.

In addition to the shallow embeddings, we train GNNs (without labels) to compute unsupervised embeddings. We employ three state-of-the-art GNN architectures, namely GCNConv~\cite{kipf2017}, GraphSAGE~\cite{GraphSAGE} and GraphConv~\cite{GraphConv}. This is done using the well-known autoencoder neural network architecture\cite{kipf2016variational} (in combination with one of the mentioned GNNs). In short, this works by training the corresponding GNN to reconstruct the input graphs. After training, an embedding is extracted from the last layer of the corresponding GNN.

\subsubsection{Supervised embeddings.} For supervised representation learning (embedding), we employ three state-of-the-art architectures, namely GCNConv \cite{kipf2017}, GraphSAGE \cite{GraphSAGE} and GraphConv \cite{GraphConv}. Given this embedding, the active and passive learning is performed using the regression model introduced in Section \ref{section:regression-model}. The reasons for this is to be consistent with the unsupervised embeddings (that will use the same regression model) and because the performance turned out to be slightly better compared to the predictions made by the last (linear) layer of the GNN. 





\subsubsection{Manual embedding} 
\begin{figure}
    \centering
    \includegraphics[width=0.8\textwidth]{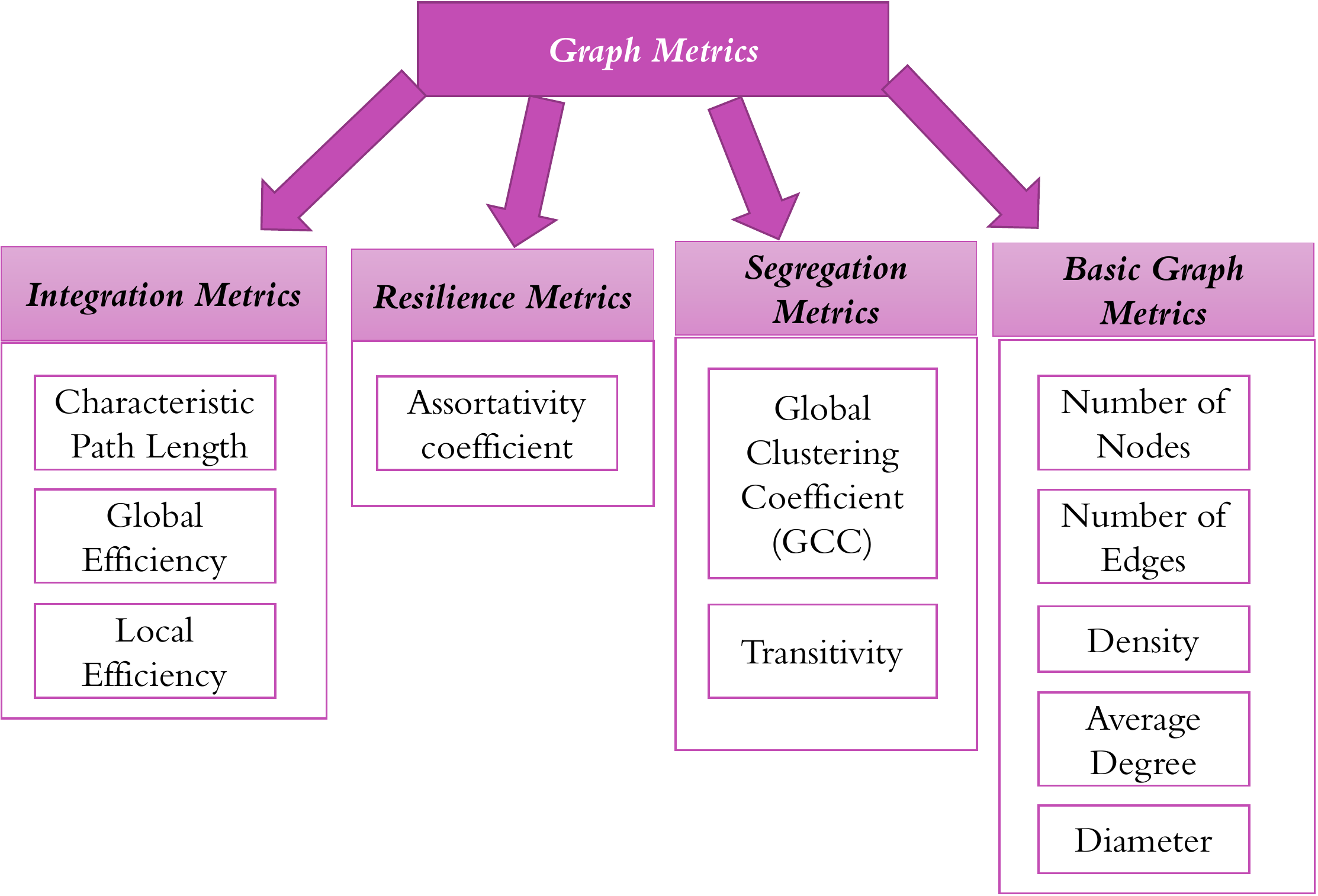}
    \caption{Hierarchy of graph-based metrics. }
    \label{fig:metrics}
\end{figure}

We also consider a manually constructed embedding by extracting a set of graph metrics for each of the graphs (data items). Figure~\ref{fig:metrics} shows a categorization of the extracted metrics. 
Below we list and explain each of the metrics.

\begin{enumerate}
    \item \textbf{Integration Metrics}\cite{PhysRevLett2001}: those metrics capture the spreading of information within the network. In particular:
\begin{itemize}
    \item \textit{Characteristic Path Length}: This metric represents the average shortest path length between all pairs of nodes in the graph. 
    \item \textit{Global Efficiency}: It measures the average inverse shortest path length between all pairs of nodes in the graph. 
    \item \textit{Local Efficiency}: Local efficiency is computed for each node as the global efficiency of its neighbourhood subgraph and then averaged over all nodes. 
\end{itemize}
\item {\textbf{Resilience Metrics}~\cite{Newman_2002}}: These metrics assess the robustness of a graph and its ability to maintain its structure and functionality despite changes or failures. In particular, we consider
\begin{itemize}
    \item \textit{Assortativity Coefficient}: this metric measures the correlation between the degrees of a node and its neighbourhood.
\end{itemize}
\item {\textbf{Segregation Metrics}\cite{PhysRevLett2001}}: they quantify the degree to which nodes in a graph tend to form tightly knit communities or clusters. Two metrics related to this category are listed below.

\begin{itemize}
    \item \textit{Global Clustering Coefficient (GCC)}\cite{watts1998collective}: it is the number of closed triplets over the total number of triplets.
    $$
    GCC = \frac{1}{n}\sum_{v \in G} \frac{2 T(v)}{deg(v)(deg(v)-1)}
    $$
    where $T(v)$ is the number of triangles through node $v$.
    \item \textit{Transitivity}: defined as
    $
        3\frac{\#triangles}{\#triads}.
    $
\end{itemize}
\item {\textit{Basic Graph Metrics}}: Basic graph metrics describe a graph's fundamental structure, size, and connectivity. In this category, we are inspired by Newman et al.~\cite{Newman2010}. Five related metrics related to this category are listed below as the following:

\begin{itemize}
    \item \textit{Number of Nodes}: The total number of nodes in the graph.
    \item \textit{Number of Edges}: The total number of edges in the graph.
    \item \textit{Diameter}: The diameter $D$ is the shortest path length between the two most distant nodes in the network.
    \item \textit{Edge Density}: The ratio of the actual number of edges to the maximum possible number of edges.
    \item \textit{Average Degree}: The average number of degrees.
\end{itemize}
\end{enumerate}

By considering these categories and their associated metrics, we can understand the graph's properties comprehensively, which can be valuable in various graph analysis and machine learning tasks.

\subsection{Incorporating Different Information}

When constructing the embeddings and performing the active/passive learning procedure outlined in Section \ref{section:al-procedure}, one can utilize different levels of information about the datasets. We describe how this is done for active learning and passive learning below. Let $\boldsymbol{X}_{\mathcal{A}}$ and $\boldsymbol{Y}_{\mathcal{A}}$ refer to the feature vectors and labels respectively of some generic dataset $\mathcal{A}$.

\subsubsection{Active Learning} For active learning we have three datasets: $\mathcal{L}_i$, $\mathcal{U}_i$ and $\mathcal{T}$. In principle, the information that can be used to construct the embeddings and perform the active learning are the labels and features of these datasets, i.e., $\boldsymbol{X}_{\mathcal{L}_i}$, $\boldsymbol{X}_{\mathcal{U}_i}$, $\boldsymbol{X}_{\mathcal{T}}$, $\boldsymbol{Y}_{\mathcal{L}_i}$, $\boldsymbol{Y}_{\mathcal{U}_i}$ and $\boldsymbol{Y}_{\mathcal{T}}$. As suggested by, for example, Munjal et al.~\cite{munjal2022robust}, it is important to separate the reported active learning results depending on what information is used. For example, if one notices improved performance when using the features of the unlabeled data items $\boldsymbol{X}_{\mathcal{U}_i}$ (through, e.g., semi-supervised learning) compared to not doing so, it is important not to fully credit this improvement to the query strategy used. Partial credit must be given to the learning algorithm used since it was able to effectively use the additional information. Note that for the active learning pipeline followed in this paper, both the construction of the embedding and the active learning can utilize different levels of information (separately). For simplicity, the active learning (given some embedding) is always done based on the training features and training labels only (i.e., supervised training based on $\boldsymbol{X}_{\mathcal{L}_i}$ and $\boldsymbol{Y}_{\mathcal{L}_i}$). However, for the construction of the embeddings, we considered four different levels of information, each of which are listed and explained below. Note that we never use $\boldsymbol{Y}_{\mathcal{T}}$, i.e., the labels of the test dataset.

\begin{itemize}[leftmargin=*]
    \item $\boldsymbol{X}_{\mathcal{L}_i}$, $\boldsymbol{X}_{\mathcal{U}_i}$ and $\boldsymbol{X}_{\mathcal{T}}$. This category is only applicable to the unsupervised embeddings (since the labels are not used). In this case, we simply construct the embedding using all available features and then embed $\mathcal{L}_i$, and $\mathcal{U}_i$ and $\mathcal{T}$ into the resulting latent space before performing the active learning.

    \item $\boldsymbol{X}_{\mathcal{L}_i}$ and $\boldsymbol{X}_{\mathcal{U}_i}$. This category is only applicable to the unsupervised embeddings (since the labels are not used). For the GNN based unsupervised embeddings it is straightforward. One begins by constructing an embedding using $\boldsymbol{X}_{\mathcal{L}_i}$ and $\boldsymbol{X}_{\mathcal{U}_i}$. Given the embedding, $\mathcal{L}_i$, $\mathcal{U}_i$ and $\mathcal{T}$ can be projected into the resulting latent space before doing the active learning. For the shallow embeddings this does not work since it is not possible to project new data items into the resulting latent space (i.e., only the data items that were used to construct the latent space can be accessed in the resulting latent space). Instead, we first construct an embedding based on $\boldsymbol{X}_{\mathcal{L}_i}$ and $\boldsymbol{X}_{\mathcal{U}_i}$ and access $\mathcal{L}_i$ and $\mathcal{U}_i$ in the resulting latent space. Then, we construct an embedding based on $\mathcal{L}_i$, $\mathcal{U}_i$ and $\mathcal{T}$ and access $\mathcal{T}$ in the resulting latent space. It should be noted that in this case $\mathcal{T}$ is in a different (but hopefully similar) feature space compared to $\mathcal{L}_i$ and $\mathcal{U}_i$. Finally, we consider the manual embedding to belong to this category since it does not use the test features when it is constructed. However, it should be noted that it is not strictly the same, since for the manual embedding the feature representation of each graph is only based on information in the graph itself (i.e., it is independent of all other graphs).

    \item $\boldsymbol{X}_{\mathcal{L}_i}$ and $\boldsymbol{Y}_{\mathcal{L}_i}$. This category is only applicable to the supervised embedding approach (based on GNNs) since it uses the labels of $\mathcal{L}_i$. In this setting one simply performs supervised training of a GNN based on $\boldsymbol{X}_{\mathcal{L}_i}$ and $\boldsymbol{Y}_{\mathcal{L}_i}$. Then, all data is projected into the latent space of the last layer of the GNN before performing the active learning.

    \item $\boldsymbol{X}_{\mathcal{L}_i}$, $\boldsymbol{Y}_{\mathcal{L}_i}$ and $\boldsymbol{X}_{\mathcal{U}_i}$. This category is only applicable to the supervised embedding approach (based on GNNs) since it uses the labels of $\mathcal{L}_i$. This setting works identically to the previous category except that we also use pseudo-labels for data items in $\mathcal{T}$ (i.e., semi-supervised learning). After some investigation, this category turned out to not lead to improved performance for our datasets and models, and is therefore not reported in the results.
\end{itemize}

\subsubsection{Passive Learning} The passive learning is conducted in a corresponding fashion to the active learning described above by simply setting $\mathcal{L} = \mathcal{L}_0$ and $\mathcal{U}_0 = \emptyset$.

\subsection{Regression model} \label{section:regression-model}

Given some graph embedding, we require a regression model to make predictions (either for passive learning or active learning). Our framework is generic enough to utilize any regression model. In this project, we investigate Gaussian Process Regressors (GPR). The reason is that GPRs are both powerful regressors while also providing an explicit uncertainty model due to their probabilistic nature \cite{GP}. This uncertainty model allows us to define a natural acquisition function that can be used in an active learning setting. This is discussed more in the next section.

\subsection{Query Strategies for Active Learning} \label{section:active-learning-strategy}

In this paper, we investigate four different query strategies (acquisition functions) in active learning experiments.

The first query strategy is to select a batch of $\mathcal{B} \subseteq \mathcal{U}_i$ samples uniformly at random, which is a common baseline strategy.

The second query strategy is based on Coreset selection and was originally introduced by Sener and Savarese~\cite{sener2018active}. It intuitively aims to select a batch $\mathcal{B} \subseteq \mathcal{U}_i$ that is maximally representative of $\mathcal{U}_i$ while simultaneously being maximally different from the samples in $\mathcal{L}_i$ (i.e., informative). We utilize the efficient k-Center-Greedy algorithm described in the original work~\cite{sener2018active}.

The third query strategy is based on the uncertainty estimations provided by the GPR. Due to the probabilistic nature of GPRs, it can produce an estimate of the variance for every data item. Let $\sigma(\boldsymbol{x})$ correspond to the variance of some data item $\boldsymbol{x} \in \mathcal{U}_i$ (estimated by the GPR). The data item in $\mathcal{U}_i$ with the highest variance (i.e., most uncertain) can then be selected according to

\begin{equation} \label{eq:gpal}
    \boldsymbol{x}^{\ast} = \operatorname*{argmax}_{\boldsymbol{x} \in \mathcal{U}_i} \sigma(\boldsymbol{x}).
\end{equation}

A batch $\mathcal{B} \subseteq \mathcal{U}_i$ can be selected by selecting the top $|\mathcal{B}|$ data items from \eqref{eq:gpal}.

Finally, we also investigate the query-by-committee (QBC) selection strategy \cite{settles.tr09}. In general, this corresponds to fitting $n$ estimators to (potentially bootstrapped) subsets of the labeled data. Then, a prediction is made by each of the estimators for all the data items in $\mathcal{U}_i$. One can then select the batch $\mathcal{B} \subseteq \mathcal{U}_i$ of data items for which the estimators disagree the most (i.e., the most uncertain ones). For example, one can compute the variance of the predictions from each of the estimators to measure this disagreement. In this paper, we employ QBC with 10 GPR estimators, each trained on a bootstrapped subset of $\mathcal{L}_i$.

\section{Experiments}
 In this section, we describe the experiments and present the results.
Each experiment has been executed on a computer with four GPU NVIDIA Tesla A40 with 48GB of memory, two CPU Xeon(R) Gold 6338, and DDR4 RAM of 256GB. The source code will be made available upon acceptance.

\subsection{Dataset Collection }

In our experiments, to increase reliability, we use two different real-world datasets of performance measurements. The first dataset (\emph{OSSBuild}) is real build data collected from the continuous integration systems of four open-source systems.
The second (\emph{HadoopTests}) is a larger dataset we have collected ourselves by repeatedly executing the unit tests of the Hadoop open-source system in a controlled environment. A summary of both datasets is provided in Table~\ref{tab:dataProjects}.
In the following subsections, we provide some additional information about each of the two datasets that we used in the experimental studies.

\subsubsection{OSSBuild Dataset}
In this dataset (originally used in Samoaa et al.~\cite{SamoaaTEPGNN}), information about test execution times in production build systems was collected for four open-source projects: systemDS, H2, Dubbo, and RDF4J. All four projects use public continuous integration servers containing (public) information about the project's builds, which we harvested for test execution times as a proxy of performance in summer 2021. Basic statistics about the projects in this dataset are described in Table~\ref{tab:dataProjects} (top). "Files" refers to the number of unit test files we collected execution times for, "Runs" is the (total) number of executions of files we extracted data for, whereas "Nodes" and "Vocabulary Size" indicate the resulting graphs (for both file and system-level parsing). 
Prior to parsing the test files, we remove code comments to reduce the number of nodes in each graph (by construction irrelevant). We note that we have 60514 more nodes for system-level parsing and 493 new vocabs. 


\newcommand{\STAB}[1]{\begin{tabular}{@{}c@{}}#1\end{tabular}}

\begin{table}[t!]
\caption{Overview of the OSSBuilds and HadoopTests datasets.}
\label{tab:dataProjects}
\centering
\scriptsize
\renewcommand{\arraystretch}{1.5}
\begin{tabular}{@{}p{0.9cm}p{1.3cm}p{3.2cm}p{0.9cm}p{0.9cm}|p{1.4cm}p{0.9cm}|p{1.4cm}p{0.9cm}@{}}
 & \textbf{Project} & \textbf{Description} & \textbf{Files} & \textbf{Runs} & \multicolumn{2}{c|}{\textbf{File-Level Parsing}} & \multicolumn{2}{c}{\textbf{System-Level Parsing}} \\ \cline{6-9}
 & & & & & \textbf{Nodes} & \textbf{Vocab.} & \textbf{Nodes} & \textbf{Vocab.} \\
\hline
\multirow{8}{*}{\rotatebox[origin=c]{90}{\textbf{OSSBuilds}}} 
& systemDS & Apache Machine Learning system for data science lifecycle & 127 & 1321 & 110651 & 3161 & 114904 & 3205 \\
& H2 & Java SQL DB & 194 & 1391 & 405706 & 17972 & 432375 & 18326 \\
& Dubbo & Apache Remote Procedure Call framework & 123 & 524 & 75787 & 4499 & 77142 & 4505 \\
& RDF4J & Scalable RDF processing & 478 & 1055 & 214436 & 10755 & 242673 & 10844 \\
& \textbf{Total} & & \textbf{922} & \textbf{4291} & \textbf{806580} & \textbf{36387} & \textbf{867094} & \textbf{36880} \\
\hline
\multirow{3}{*}{\rotatebox[origin=c]{90}{\textbf{HadoopTests}}} 
& \textbf{Hadoop} & Apache framework for processing large datasets on clusters & \textbf{2895} & \textbf{24348} & \textbf{4314360} & \textbf{135408} & \textbf{5090798} & \textbf{138952} \\ [1.7cm]
\hline
\end{tabular}
\end{table}

\subsubsection{HadoopTests Dataset}
To address limitations with the OSSBuilds dataset (primarily the limited number of files for each individual project in the dataset), we additionally collected a second dataset for this study. We selected the Apache Hadoop framework since it entails a large number of test files (2895) of sufficient complexity. We then executed all unit tests in the project five times, recording the execution duration of each test file as reported by the JUnit framework (in millisecond granularity). As an execution environment for this data collection, we used a dedicated virtual machine running in a private cloud environment, with two virtualized CPUs and 8 GByte of RAM. Following performance engineering best practices, we deactivated all other non-essential services while running the tests.
Statistics about the HadoopTests dataset are described in Table~\ref{tab:dataProjects} (bottom).


Since we have more files in HadoopTests, we have more added nodes to the system-level parsing setting. Thus 776438 nodes are added to the graphs in the system-level parsing, and we get 3544 more vocabs. 

\subsection{Analysis of Graphs} \label{section:GraphExp}
We want to annotate each source code file with the corresponding scalar value related to execution time. The source code is represented as a graph. In particular, each graph represents a Java source code file (a JUnit test case). As aforementioned, the base structure is a tree that is then extended to a graph adding edges representing program control flow~\cite{SamoaaTEPGNN}.

Table \ref{tab:graphmetrics} shows the average statistics of the input graphs. In particular, we report the average number of nodes ($|V|$), the average number of edges ($|E|$), the density, the average global clustering coefficient ($GCC$), the average number of cycles and the average tree similarity.
We define a simple function to measure, how similar the graph is to a tree ($tree-sim$) as the number of edges that have to be removed to convert the graph into a tree, i.e., 
\begin{equation}\label{eq:treesim}
    tree-sim = \dfrac{|E| - (|V|-1)}{(|V|-1)(\dfrac{|V|}{2}-1)} \, .
\end{equation}

The formula has to be interpreted as the number of edges of the graphs minus the number of edges of a tree with $N$ nodes, normalized. If the input graph is a tree, then we have that $tree-sim$ is equal to 0, while if the graph is complete, $tree-sim$ is equal to 1.

\begin{table}[h!]
    \caption{Average statistics of the input graphs of System Level Parsing.}
    \label{tab:graphmetrics}
    \centering
    \begin{tabular}{l|cccccccc}
         \hline
         Dataset & type & $|V|$ & $|E|$ & Diameter & Density & $GCC$ & $tree-sim$ \\
         \hline
         \multirow{2}{*}{\textbf{OSSBuilds}}&File-level   & 875 & 1679  & 14 & 0.014 & 0.16 & 0.007 \\ 
                                            &System-level & 940 & 1848  & 13 & 0.013 & 0.15 & 0.006 \\ 
         \hline
         \multirow{2}{*}{\textbf{HadoopTests}}&File-level & 1490 & 1848 & 15 & 0.005 & 0.15 & 0.003 \\ 
                                            &System-level & 1734 & 3428 & 14 & 0.006 & 0.15 & 0.003 \\ 
         \hline
         \hline
    \end{tabular}
\end{table}

From Table \ref{tab:graphmetrics}, it is easy to see that the input graph has a high diameter. In fact, if we generate a random graph \cite{batagelj2005efficient} with the same number of nodes, and the same density as the original ones, we obtain an average diameter of 2 and 4 for OSSBuilds and HadoopTests, respectively.
It is also easy to see that the input graphs are quite sparse. Finally, in both datasets, the $tree-sim$ is close to zero, thus we can conclude that input graphs are similar to trees.
We report a detailed analysis of the input graphs in appendix \ref{appendix:graphanal}.

\subsection{Results}
In this section, we present the results of both the passive and active learning experiments, utilizing the Scikit-learn \cite{scikit-learn} implementation of Gaussian Process Regressors with a Matern kernel. In the passive setting, the hyperparameters of the Matern kernel were fine-tuned. For active learning, the hyperparameters of the Matern kernel were fine-tuned in each iteration based on the currently available labeled data in $\mathcal{L}_i$. The GNN models used for both supervised and unsupervised embeddings consist of three layers with 30 neurons each. Since each layer learns a node representation, we compute the graph representation by concatenating the sum, average, and max of the node representation, resulting in an embedding of 90 dimensions. The Adam optimizer \cite{kingma2014adam} is employed with a learning rate of $0.001$, and the loss used is the Mean Squared Error.


\subsubsection{Passive Learning }
\label{sec:passivelearning}
To perform passive learning, we utilize a training set $\mathcal{L}$ and a test set $\mathcal{T}$. For each embedding, we train a Gaussian process (GP) using $\mathcal{L}$ and then use it to predict the execution time of all test data items in $\mathcal{T}$. We measure the quality of the predictions by computing the Pearson correlation score. \textcolor{black}{Additionally, all passive learning results correspond to the average of 15 runs with different seeds, where for each method, the mean and standard deviation (STD) values are reported.
}

We will show the results for system-level parsing and file-level parsing. 
\textcolor{black}{\paragraph{\textbf{System-Level Parsing}}
Table~\ref{tab:resultsunsup} displays the results for the unsupervised embeddings based on both train and test features, as well as only train features for GNN since it is not doable for shallow embedding methods. The shallow embedding methods are unsupervised techniques used for learning node representations in graph-structured data. Thus, when we are using these unsupervised approaches, we typically compute the embeddings on the entire dataset, which includes both the training and test features (without using the target variables of test data). The rationale behind this is that these embedding techniques learn representations based on the graph's structural properties and node connections rather than being tailored to specific labels or tasks. This means that the embeddings are primarily influenced by the relationships and interactions between nodes in the graph. Training the embedding model on the training set alone, the model would not have access to information from the test set. As a result, the embeddings learned from the training set might not adequately capture the patterns or relationships present in the test set. This could lead to suboptimal performance when using these embeddings for our downstream task. That's why the results in Table~\ref{tab:resultsunsup} when test features are not used are NA (i.e., not applicable). However, to demonstrate this point, we add additional results in the appendix~\ref{append_results} for the shallow embedding when we have two different feature spaces for embeddings because we train the embedding model on the train set, then train the same model again on the test set. The previous delima is not applicable for GNN since we can train the GNN based on a training set to compute the embedding and use the trained GNN model to map the test data and infer the embedding for the test set.}

\begin{table}[h!]
\caption{Results for Unsupervised Embedding for graphs of System Level Parsing.}
\label{tab:resultsunsup}
\centering
\begin{tabular}{@{}c|cc|cc|cc@{}}
 & & & \multicolumn{2}{c|}{Train and Test Features} & \multicolumn{2}{c}{Train Features} \\
 \hline
 & & & OSSBuilds & HadoopTests & OSSBuilds & HadoopTests \\
\hline
 \multirow{9}{*}{\STAB{\rotatebox[origin=c]{-90}{Shallow Embedding}}} & \multicolumn{2}{c|}{Graph2Vec} & $\textbf{0.73} \pm 0.03$ & $\textbf{0.75} \pm 0.02$  & {NA} & {NA} \\
& \multirow{2}{*}{GR} & mean & $0.45 \pm 0.04$  & $0.47 \pm 0.02$ & NA & NA \\
& & sum & $0.40 \pm 0.05$ & $0.43 \pm 0.03$  & NA &NA \\
& \multirow{2}{*}{HOPE} & mean & $0.19 \pm 0.07$  & $0.06 \pm 0.03$ & NA & NA \\
& & sum & $0.20 \pm 0.08$ & $0.35 \pm 0.04$ & NA & NA \\
& \multirow{2}{*}{DeepWalks} & mean & $0.37 \pm 0.06$ & $0.44 \pm 0.02$ & NA & NA \\
& & sum & $0.36 \pm 0.06$ & $0.43 \pm 0.04$ & NA & NA \\
& \multirow{2}{*}{Node2Vec} & mean & $0.33 \pm 0.06$ & $0.42 \pm 0.03$ & NA & NA \\
& & sum & $0.36 \pm 0.06$ & $0.42 \pm 0.04$ & NA & NA \\
\hline
\multirow{3}{*}{GNN} & \multicolumn{2}{c|}{GCNConv} & $0.41 \pm 0.06$ & $0.48 \pm 0.03$ & $0.44 \pm 0.05$ & $0.48 \pm 0.0$3 \\
& \multicolumn{2}{c|}{GraphSAGE} & $0.37 \pm 0.06$ & $0.42 \pm 0.04$ & $0.38 \pm 0.04$ & $0.45 \pm 0.05$ \\
& \multicolumn{2}{c|}{GraphConv} & $0.43 \pm 0.06$ & $0.49 \pm 0.03$ & $0.44 \pm 0.07$ & $0.49 \pm 0.03$ \\
\end{tabular}
\end{table}

Thus looking at the results of Tables~\ref{tab:resultsunsup}, we notice that Graph2Vec attains the highest scores of 0.73 and 0.75 for the OSSBuilds and HadoopTests datasets, respectively. \textcolor{black}{With  Graph2Vec, the entire graph is embedded, but when it comes to other shallow embedding methods, the embedding is for node level. Thus, in order to have the embedding for the entire graph, the embedding is aggregated using \textit{mean} and \textit{sum} aggregation functions. GR (for both datasets), DeepWalks, and Node2Vec (for the HadoopTests dataset) with both aggregation functions achieve a reasonable Pearson correlation score. On the other hand, HOPE is the worst. The results of shallow embeddings are more stable for Hadooptests since the STD is in the range of [0.02,0.04], which is not the case for OSSBuilds when the STD range is [0.03,0.08]. This is reasonable because by looking at Table~\ref{tab:dataProjects}, we can see that OSSBuilds contains four different projects for four different domains, which is not the case for HadoopTets, where all code files are related to one project.  } 

\textcolor{black}{The performance of the GNN-based methods is slightly better when the test features are not used in the embedding.
GraphConv is the best GNN model in both cases. The unsatisfactory performance of GNNs is not surprising, as unsupervised graph representation learning requires vast data.}


\begin{table}[h!]
\caption{Results for Supervised and Manual Embedding for graphs of System Level Parsing.}
\label{tab:resultsup}
\centering
\begin{tabular}{@{}c|cc|cc@{}}
& & & \multicolumn{2}{c|}{Train Features} \\
\hline
& & & OSSBuilds & HadoopTests \\
\hline
\multirow{3}{*}{Supervised Embedding (GNN)} & \multicolumn{2}{c|}{GCNConv} & $0.59 \pm 0.04$ & $0.64 \pm 0.03$  \\
& \multicolumn{2}{c|}{GraphSAGE} & $0.61 \pm 0.04$ & $\textbf{0.67} \pm 0.02$ \\
& \multicolumn{2}{c|}{GraphConv} & $\textbf{0.65} \pm 0.04$ & $0.66 \pm 0.02$ \\
\hline
Manual Embedding & & & $0.56 \pm 0.04$ & $0.59 \pm 0.02$ \\
\end{tabular}
\end{table}

The results for the supervised embeddings based on the train features $\boldsymbol{X}_{\mathcal{L}}$ and train labels $\boldsymbol{Y}_{\mathcal{L}}$ are presented in Table~\ref{tab:resultsup}. The correlation obtained with GNNs is shown in the first rows, while the results obtained using the manual embedding are reported in the last row. It is evident from the table that the performance of the GNN-based approaches is superior to that of the manual embeddings for both datasets. \textcolor{black}{Thus, GraphConv performs the best for OSSBuilds with an average correlation score of 0.65 and STD of 0.04, whereas GraphSAGE is the highest average correlation score ( with a very tiny difference compared to other GNN methods).}

\begin{table}[h!]
\caption{Results for Unsupervised Embedding for graphs of File Level Parsing.}
\label{tab:resultsunsup_NOSUT}
\centering
\begin{tabular}{@{}c|cc|cc|cc@{}}
 & & & \multicolumn{2}{c|}{Train and Test Features} & \multicolumn{2}{c}{Train Features} \\
 \hline
 & & & OSSBuilds & HadoopTests & OSSBuilds & HadoopTests \\
\hline
 \multirow{9}{*}{\STAB{\rotatebox[origin=c]{-90}{Shallow Embedding}}} & \multicolumn{2}{c|}{Graph2Vec} & $\textbf{0.74} \pm 0.03$ & $\textbf{0.74} \pm 0.02$ & NA & {NA} \\
& \multirow{2}{*}{GR} & mean & $0.58 \pm 0.03$ & $0.50 \pm 0.03$ & NA & NA \\
& & sum & $0.47 \pm 0.05$ & $0.46 \pm 0.04$ & NA & NA \\
& \multirow{2}{*}{HOPE} & mean & $0.16 \pm 0.05$ & $0.06 \pm 0.03$ & NA & NA \\
& & sum & $0.16 \pm 0.05$ & $0.37 \pm 0.05$ & NA & NA \\
& \multirow{2}{*}{DeepWalks} & mean & $0.42 \pm 0.05$ & $0.47 \pm 0.03$ & NA & NA \\
& & sum & $0.41 \pm 0.05$ & $0.46 \pm 0.04$ & NA & NA \\
& \multirow{2}{*}{Node2Vec} & mean & $0.30 \pm 0.06$ & $0.20 \pm 0.03$ & NA & NA \\
& & sum & $0.25 \pm 0.07$ & $0.40 \pm 0.04$ & NA & NA \\
\hline
\multirow{3}{*}{GNN} & \multicolumn{2}{c|}{GCNConv} & $0.47 \pm 0.05$ & $0.52 \pm 0.04$ & $0.46 \pm 0.04$ & $0.50 \pm 0.04$ \\
& \multicolumn{2}{c|}{GraphSAGE} & $0.44 \pm 0.06$ & $0.44 \pm 0.04$ & $ 0.42 \pm 0.04$ & $0.42 \pm 0.04$ \\
& \multicolumn{2}{c|}{GraphConv} & $0.48 \pm 0.05$ & $0.51 \pm 0.04$ & $ 0.47 \pm 0.05$ & $ 0.51 \pm 0.03$ \\
\end{tabular}
\end{table}

\paragraph{\textbf{File-Level Parsing}}
Table~\ref{tab:resultsunsup_NOSUT} displays the results for the unsupervised embeddings based on both train and test features and only train features for GNN for File-Level parsing. When utilizing both the training and testing features, we notice consistent results in the average scores of Graph2Vec for both datasets—0.74 each. This convergence aligns with the score we obtained in System-Level Parsing. \textcolor{black}{ In this experiment setting, GR for both datasets is significantly improved, whereas DeepWalks maintains the same score. Unlike the system-level parsing setting, the embedding quality for Node2Vec decreased and the opposite for GR. That explains why Node2Vec performs better with more nodes and edges graph data. So, the gap between GR (which remains the second best performing unsupervised embedding approaches) and the other two approaches (DeepWalks and Node2vec) is extended more. Moreover, HOPE remains the worst-performing approach regarding embedding quality. 
As for GNN in this setting, the average score is significantly improved when test features are used in embedding compared to the System-Level Parsing setting. More interestingly, the aforementioned results are better than the ones when the test features are not used. 
Nevertheless, GraphConv is the best GNN model for HadoopTests and OSSBuilds } 


\begin{table}[h!]
\caption{Results for Supervised and Manual Embedding for graphs of File Level Parsing.}
\label{tab:resultsup_NOSUT}
\centering
\begin{tabular}{@{}c|cc|cc@{}}
& & & \multicolumn{2}{c|}{} \\
\hline
& & & OSSBuilds & HadoopTests \\
\hline
\multirow{3}{*}{Supervised Embedding (GNN)} & \multicolumn{2}{c|}{GCNConv} & $0.61 \pm 0.04$ & $0.66 \pm 0.02$ \\
& \multicolumn{2}{c|}{GraphSAGE} & ${0.64} \pm 0.03$ & $\textbf{0.68} \pm 0.02$ \\
& \multicolumn{2}{c|}{GraphConv} & $\textbf{0.67} \pm 0.02$ & $\textbf{0.68} \pm 0.01$ \\
\hline
Manual Embedding & & & $0.64 \pm 0.05$ & $0.61 \pm 0.02$ \\
\end{tabular}
\end{table}

\textcolor{black}{ As for supervised results in Table~\ref{tab:resultsup_NOSUT}, the results for all GNN-based models are a bit improved compared to Syetem-Level Parsing. The same is true regarding Manual Embedding. The reason for this improvement might be that we have fewer nodes and edges with File-level parsing, which means less sparsity. }

 Overall, for passive learning, Graph2Vec with test features achieves the best score for both datasets and settings. The reason why Graph2Vec performs well could be because our input graphs are similar to trees (see Section \ref{section:GraphExp}). In fact, Graph2Vec explores a much deeper path within the input graph compared to GNN. On the other hand, GNNs in a supervised setting deliver reasonable results for both datasets (unlike the unsupervised GNN embedding). This is likely because the labels are utilized. The manual embedding also yields an acceptable score compared to the shallow embeddings (except Graph2Vec).

\subsubsection{Active Learning} 
Given an embedding, the active learning experiments were conducted as outlined in Section \ref{section:al-procedure}. We investigate different sizes of the initially labeled dataset $|\mathcal{L}_0|$ and the batch size $|\mathcal{B}|$. Additionally, all active learning results correspond to the average of 15 runs with different seeds, where the variance of the runs is indicated by a shaded color.

\begin{figure}[htb!] 
\centering
\subfigure[]{\label{subfig:he1}\includegraphics[width=0.24\linewidth]{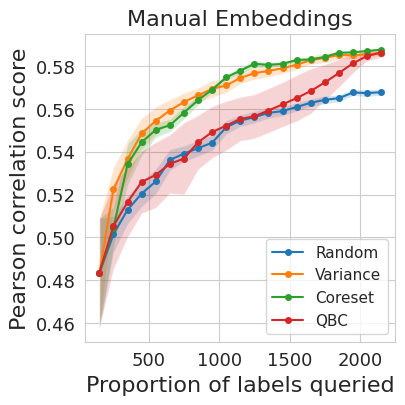}}
\subfigure[]{\label{subfig:he3}\includegraphics[width=0.24\linewidth]{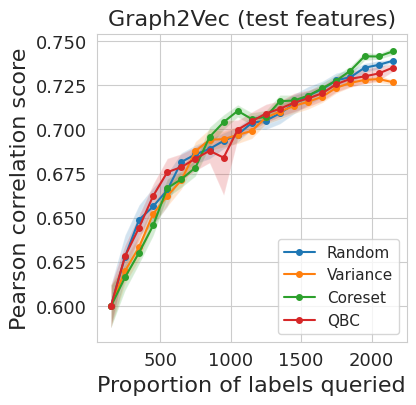}}
\subfigure[]
{\label{subfig:he4}\includegraphics[width=0.24\linewidth]{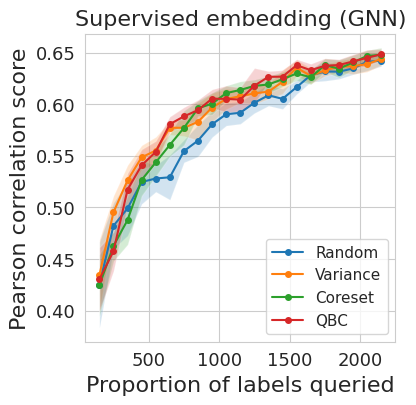}}
\caption{Active learning results for all embeddings for the HadoopTests dataset (System Level Parsing) with $|\mathcal{L}_0| = 150$ and $|\mathcal{B}| = 100$.}
\label{fig:hadoop-embeddings}
\end{figure}

The active learning experiments investigate three different graph embeddings (based on the passive learning results): manual embedding,
Graph2Vec (with test features) and the supervised (GNN) embedding. For each embedding, we use the four query strategies outlined in Section \ref{section:active-learning-strategy} (i.e., coreset, variance, QBC, and random).
\paragraph{\textbf{System Level Parsing Graphs:}} 
In this section, we present the results of the experiments conducted on system-level parsing. 
Figure \ref{fig:hadoop-embeddings} shows one plot for each of the three embeddings for the HadoopTests dataset. In each plot, the results for each query strategy are illustrated. For this experiment we set $|\mathcal{L}_0| = 150$ and $|\mathcal{B}| = 100$. 

\begin{figure}[htb!] 
\centering
\subfigure[]{\label{subfig:hqs1}\includegraphics[width=0.24\linewidth]{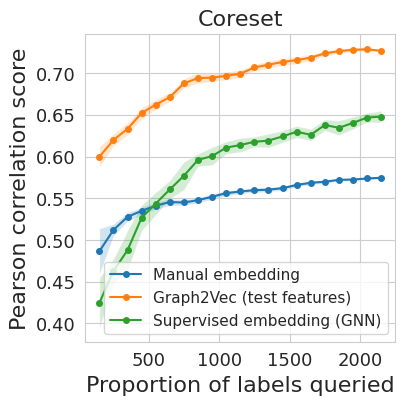}}
\subfigure[]{\label{subfig:hqs2}\includegraphics[width=0.24\linewidth]{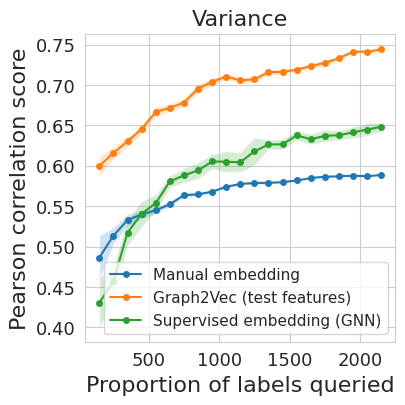}}
\subfigure[]{\label{subfig:hqs3}\includegraphics[width=0.24\linewidth]{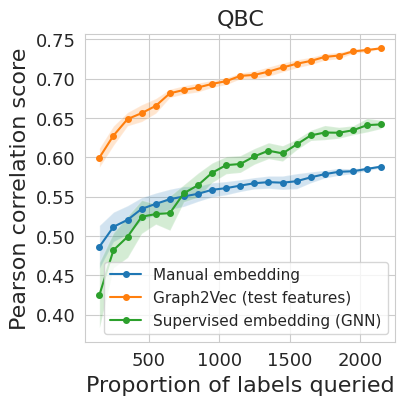}}
\subfigure[]
{\label{subfig:hqs4}\includegraphics[width=0.24\linewidth]{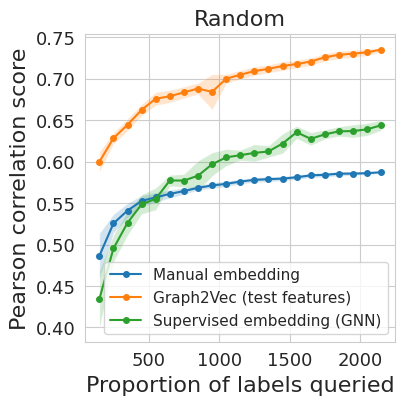}}
\caption{Active learning results for all query strategies for the HadoopTests dataset (System Level Parsing) with $|\mathcal{L}_0| = 150$ and $|\mathcal{B}| = 100$.}
\label{fig:hadoop-qs}
\end{figure}

We observe that variance is a very strong baseline for all embeddings in this setting, as it is almost the best-performing method overall. 
As for active learning strategies, variance and coreset both achieve the best results for manual embedding, whereas uniform random is the worst which is the case in supervised embedding using GNN where the other three methods are close to each other for all number of samples. 
In Contrast, when test features are involved in Graph2Vec embedding, the four query strategies are close to each other with slight privilege for coreset. 

Figure \ref{fig:hadoop-qs} shows one plot for each of the four query strategies applied to the HadoopTests dataset for the same experiment. In each plot, the results for each embedding are visualized. Graph2Vec, when the test features are used, outperforms the other two embeddings for all query strategies. The supervised embedding (GNN) that utilizes the train features and train labels performs the second best compared to Manual embedding.

\begin{figure}[htb!] 
\centering
\subfigure[]{\label{subfig:p1}\includegraphics[width=0.24\linewidth]{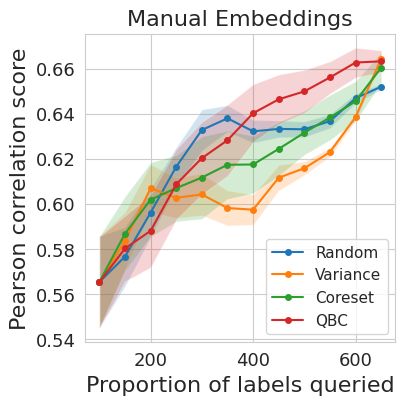}}
\subfigure[]{\label{subfig:p3}\includegraphics[width=0.24\linewidth]{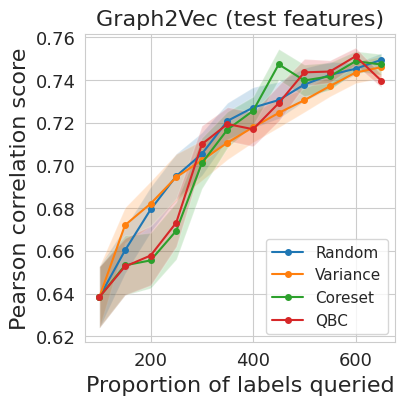}}
\subfigure[]
{\label{subfig:p4}\includegraphics[width=0.24\linewidth]{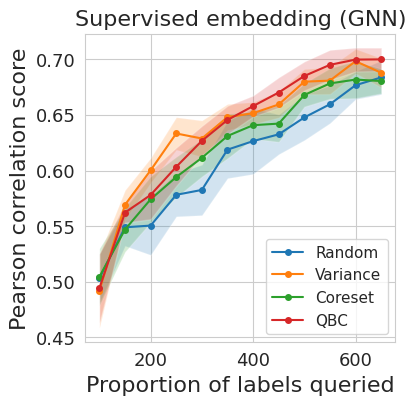}}
\caption{Active learning results for all embeddings for the OSSBuilds dataset (System Level Parsing) with $|\mathcal{L}_0| = 100$ and $|\mathcal{B}| = 50$.}
\label{fig:pro-embeddings}
\end{figure}

\begin{figure}[htb!] 
\centering
\subfigure[]{\label{subfig:pe1}\includegraphics[width=0.24\linewidth]{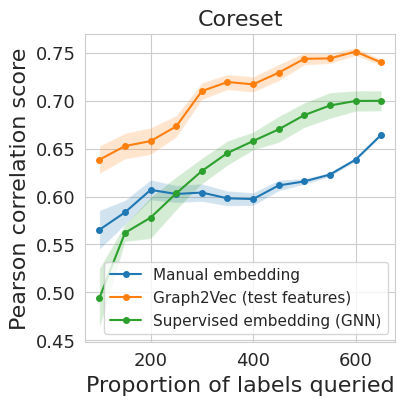}}
\subfigure[]{\label{subfig:pe2}\includegraphics[width=0.24\linewidth]{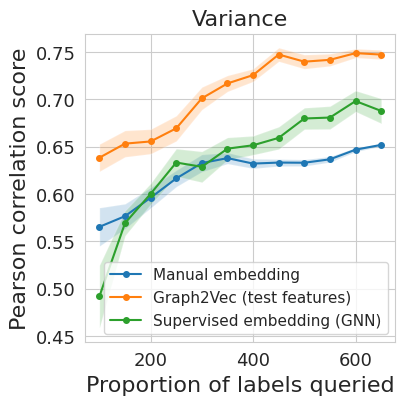}}
\subfigure[]{\label{subfig:pe3}\includegraphics[width=0.24\linewidth]{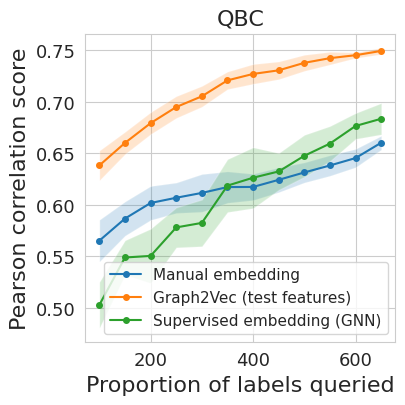}}
\subfigure[]
{\label{subfig:pe4}\includegraphics[width=0.24\linewidth]{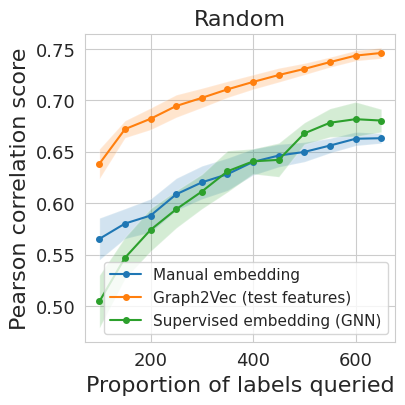}}
\caption{Active learning results for all query strategies for the OSSBuilds dataset (System Level Parsing) with $|\mathcal{L}_0| = 100$ and $|\mathcal{B}| = 50$.}
\label{fig:pro-qs}
\end{figure}

Figures \ref{fig:pro-embeddings} and \ref{fig:pro-qs} illustrate similar investigations on the OSSBuilds dataset. For this experiment we set $|\mathcal{L}_0| = 100$ and $|\mathcal{B}| = 50$. In Figure~\ref{fig:pro-embeddings}, we can see that all four selection methods are close to each other in Supervised embedding (even random is the worst). However, in manual embeddings, random performs the best when the number of samples is between 200 and 400. But when we have small initial samples, corset and variance still perform the best, whereas QBC is dominant when we have more samples. As for Graph2Vec, when test features are manipulated in embedding, Variance achieves the best results in the first 300 samples, and then the corset and QBC become the best afterwards. Regarding Figure~\ref{fig:pro-qs}, the results are somehow consistent with the similar figure for the HadoopTest dataset. Thus, Graph2Vec, when the test features are used, is the best across all selection methods, and supervised embedding using GNN is the second best.


In conclusion, we observe that active learning can lead to good performance with a small portion of the total data items labelled. In general, the Graph2Vec embedding (with test features) is superior in terms of performance, with the supervised GNN embedding in second place (which utilizes train labels). Furthermore, the adaptive query strategies (i.e., variance, coreset, and QBC) do improve on random selection, but not very significantly in some cases. The reason could be multifold. First, we are in a regression setting. Second, we are in a deep learning setting where batch selection is a requirement. Third, it can be due to the properties of the data (see Appendix~\ref{appendix:graphanal}). All of these factors are generally known to make it significantly harder to outperform the random baseline in an active learning setting. 

\paragraph{\textbf{File Level Parsing Graphs:}}

\begin{figure}[htb!] 
\centering
\subfigure[]{\label{subfig:he1_file}\includegraphics[width=0.24\linewidth]{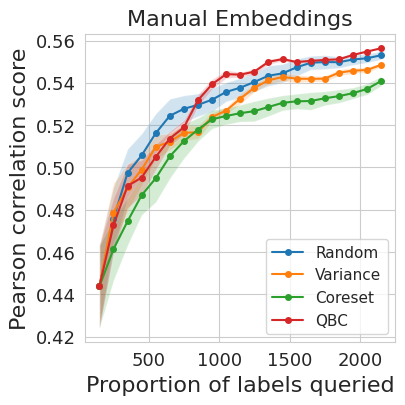}}
\subfigure[]{\label{subfig:he3_file}\includegraphics[width=0.24\linewidth]{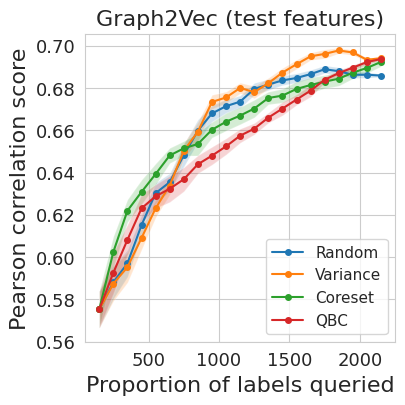}}
\subfigure[]
{\label{subfig:he4_file}\includegraphics[width=0.24\linewidth]{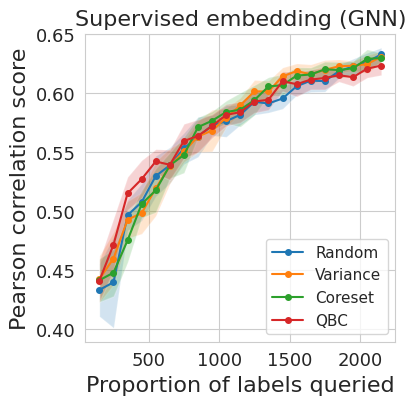}}
\caption{Active learning results for all embeddings for the HadoopTests dataset (File Level Parsing) with $|\mathcal{L}_0| = 150$ and $|\mathcal{B}| = 100$.}
\label{fig:hadoop-embeddings_NOSUT}
\end{figure}

\begin{figure}[htb!] 
\centering
\subfigure[]{\label{subfig:hqs1_file}\includegraphics[width=0.24\linewidth]{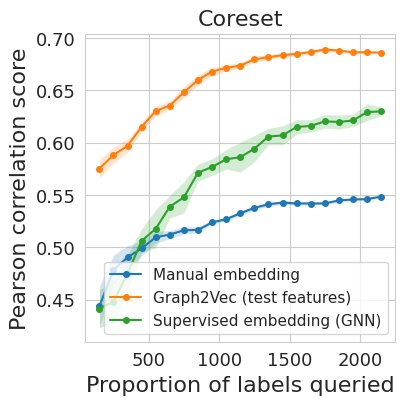}}
\subfigure[]{\label{subfig:hqs2_file}\includegraphics[width=0.24\linewidth]{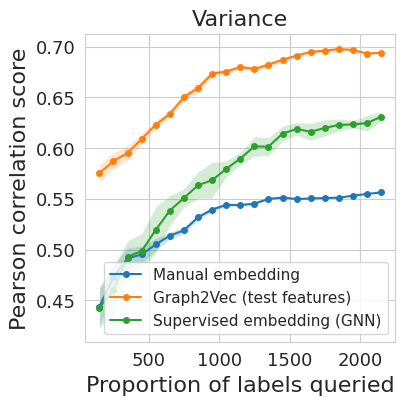}}
\subfigure[]{\label{subfig:hqs3_file}\includegraphics[width=0.24\linewidth]{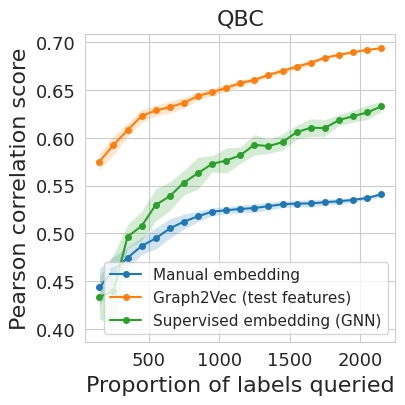}}
\subfigure[]
{\label{subfig:hqs4_file}\includegraphics[width=0.24\linewidth]{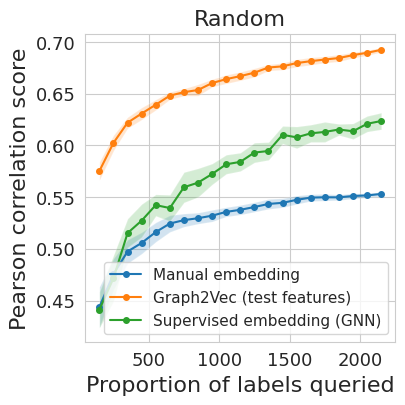}}
\caption{Active learning results for all query strategies for the HadoopTests dataset (File Level Parsing) with $|\mathcal{L}_0| = 150$ and $|\mathcal{B}| = 100$.}
\label{fig:hadoop-qs_NOSUT}
\end{figure}

In Figure~\ref{fig:hadoop-embeddings_NOSUT}, compared to Figures~\ref{fig:hadoop-embeddings}, there is no notable difference when it comes to Supervised embedding using GNN and Graph2Vec when test features are utilized. However, the clear difference is in Manual embeddings where the gap between the query strategies is noticeably reduced. 

When it comes to the four query strategies applied to HaddopTests, there is no clear difference in the results obtained in the system-level parsing setting. 

As for the Ossbuilds dataset, in Figure~\ref{fig:hadoop-embeddings_NOSUT}, there is a remarkable change compared to the system-level parsing setting. As for Manual embedding, despite QBC and random being slightly better in the initial 200 samples, coreset then becomes the dominant choice afterwards. In system-level parsing, the query strategies are competing to be the best according to the number of samples Figure~\ref{fig:hadoop-embeddings}. Thus as coreset and variance perform better in the initial 200 samples, random then perform the best in the proceeding 200 samples, and then QBC becomes the best when we have 400 samples. In Graph2vec the only difference compared with system-level parsing is that the gap between the variance and other selection methods becomes a bit higher.

\begin{figure}[htb!] 
\centering
\subfigure[]{\label{subfig:pe1_file}\includegraphics[width=0.24\linewidth]{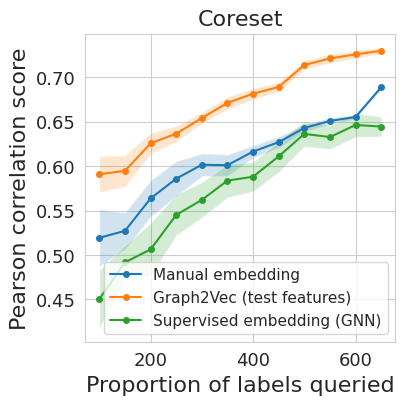}}
\subfigure[]{\label{subfig:pe2_file}\includegraphics[width=0.24\linewidth]{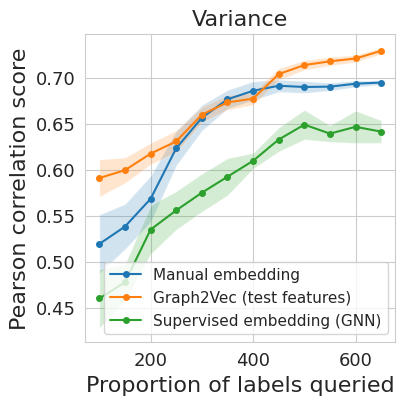}}
\subfigure[]{\label{subfig:pe3_file}\includegraphics[width=0.24\linewidth]{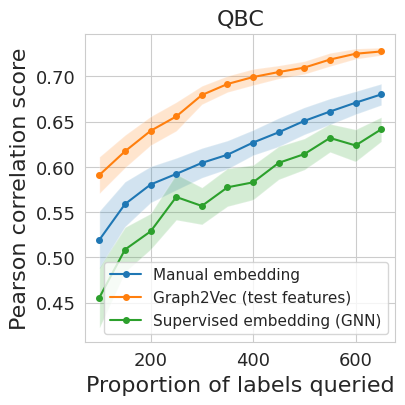}}
\subfigure[]
{\label{subfig:pe4_file}\includegraphics[width=0.24\linewidth]{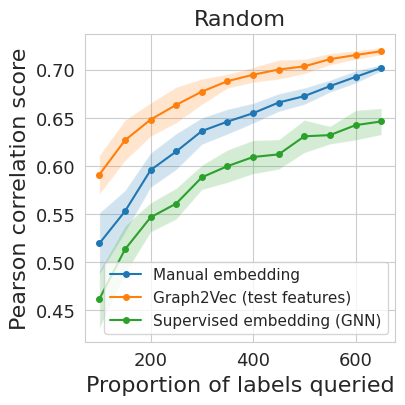}}
\caption{Active learning results for all query strategies for the OSSBuilds dataset (File Level Parsing) with $|\mathcal{L}_0| = 100$ and $|\mathcal{B}| = 50$.}
\label{fig:pro-qs_NOSUT}
\end{figure}

Regarding Figure~\ref{fig:pro-qs_NOSUT}, similar to system-level parsing, Graph2Vec when test features are used yields the best results. However, manual embedding becomes the second-best choice across all selection methods.

\begin{figure}[htb!] 
\centering
\subfigure[]{\label{subfig:p1_file}\includegraphics[width=0.24\linewidth]{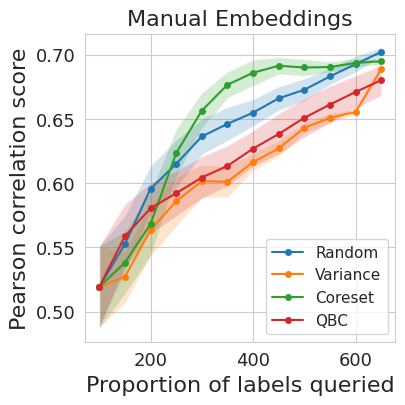}}
\subfigure[]{\label{subfig:p3_file}\includegraphics[width=0.24\linewidth]{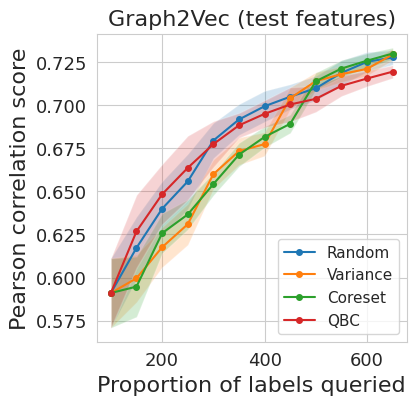}}
\subfigure[]
{\label{subfig:p4_file}\includegraphics[width=0.24\linewidth]{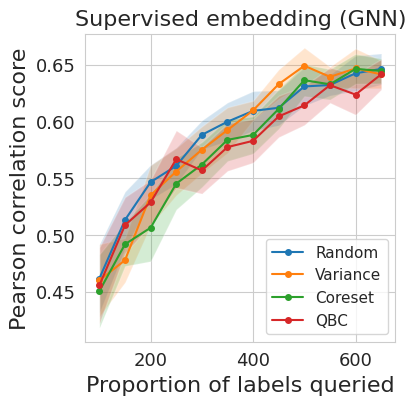}}
\caption{Active learning results for all embeddings for the OSSBuilds dataset (File Level Parsing) with $|\mathcal{L}_0| = 100$ and $|\mathcal{B}| = 50$.}
\label{fig:pro-embeddings_NOSUT}
\end{figure}

\section{Conclusion}
In this paper, we proposed a novel framework for active learning to graph data (i.e., on the graph level not to node level). Within this framework, we studied the impact of the use of additional information on active learning (namely the features of test data and the partially available queried labels for embedding graphs). 
We investigated the framework for the novel real-world application of software source code performance prediction. 
Expectedly, we observed that the performance depends strongly on the quality of the embedding. In turn, we saw that the quality of the embedding depends both on the inductive biases of the embedding algorithm used as well as what information is used when constructing it. Based on the results, Graph2Vec seems to work particularly well for our graph data. Additionally, utilizing the test features and/or the training labels leads to improved performance. 
The active learning results indicate that it is possible to achieve good performance using only a small subset of the labels. 
Our work provides a novel open-source framework for researchers to
investigate various active learning methods for graph data.

\section*{Acknowledgments}
This work received financial support from the Swedish Research Council VR under grant number 2018-04127 (Developer-Targeted Performance Engineering for Immersed Release and Software Engineers). The work of Linus Aronsson and Morteza Haghir Chehreghani was partially supported by the Wallenberg AI, Autonomous Systems and Software Program (WASP) funded by Knut and Alice Wallenberg Foundations. Finally, the computations and data handling were enabled by resources provided by the National Academic Infrastructure for Supercomputing in Sweden (NAISS) and the Swedish National Infrastructure for Computing (SNIC), partially funded by the Swedish Research Council through grant agreement no. 2022-06725 and no. 2018-05973.

\bibliographystyle{unsrt}  
\bibliography{references}  

\appendix
\newpage




\section{Graph analysis} \label{appendix:graphanal}

In this section, we do a deeper investigation of the graph topology of our dataset.

\subsection{Basic topology}

\begin{figure}[h!] 
\centering
\subfigure[]{\label{fig:app:nbnode}\includegraphics[width=\linewidth]{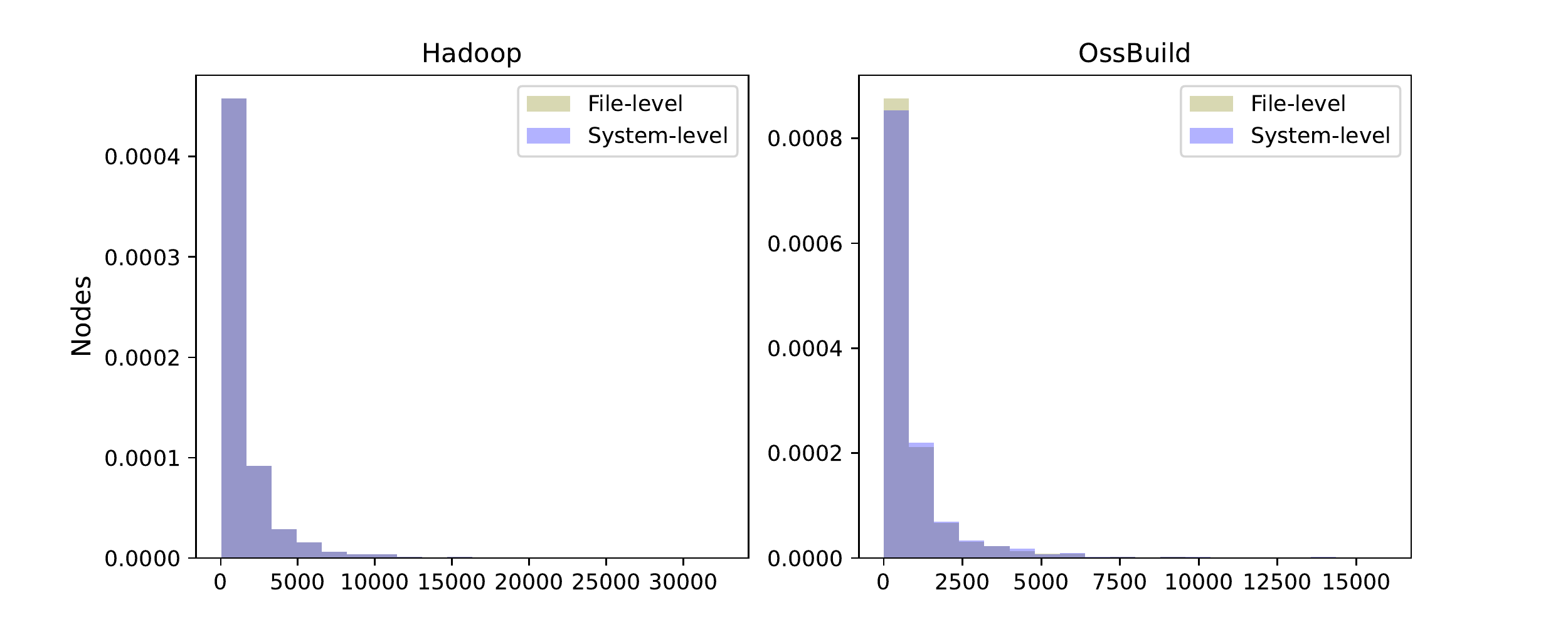}}
\subfigure[]{\label{fig:app:nbedge}\includegraphics[width=\linewidth]{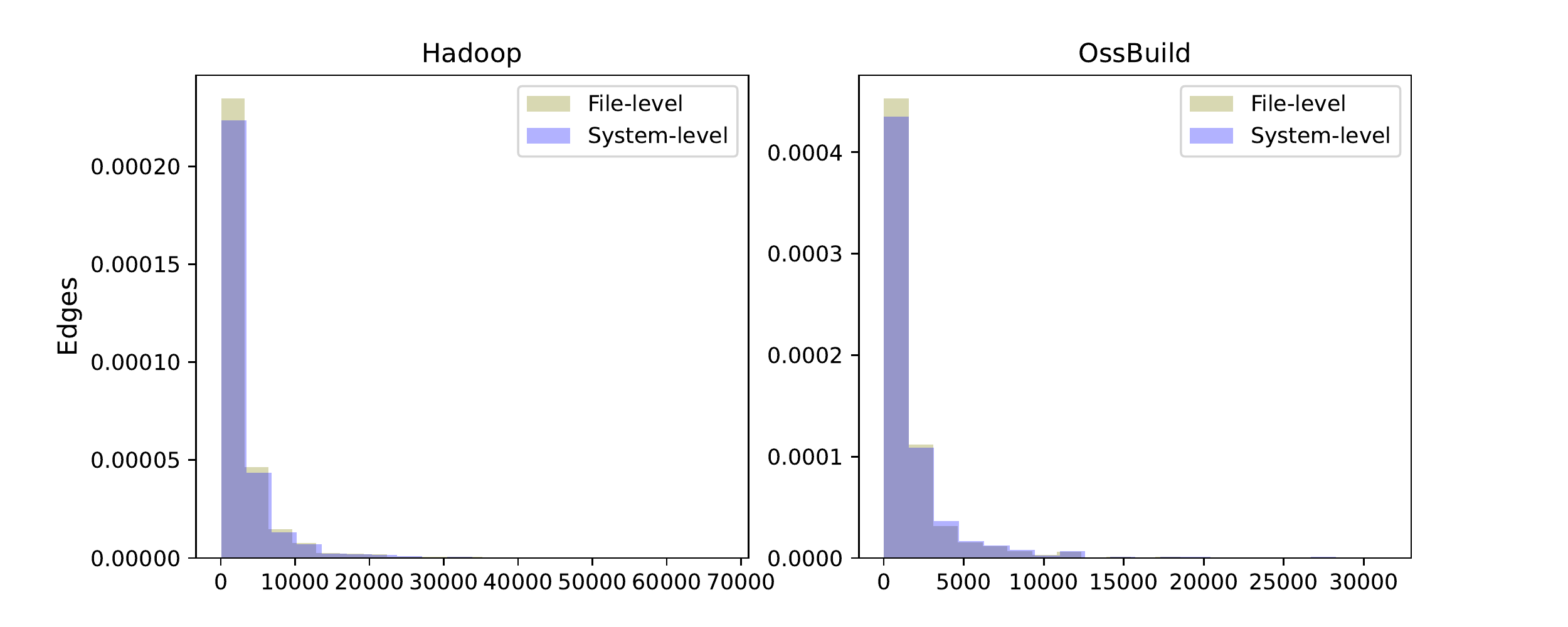}}
\caption{Distributions of the number of nodes and edges in Hadoop (left) and OssBuild (right) for both file-level and system-level settings.}
\label{fig:app:nodeedge}
\end{figure}

Figure \ref{fig:app:nodeedge} displays node (Figure \ref{fig:app:nbnode}) and edge
(Figure \ref{fig:app:nbedge}) distributions, respectively. The data indicate a minimal disparity between file and system levels in terms of both statistics.

\begin{figure}[h!]
    \centering
    \includegraphics[width=\textwidth]{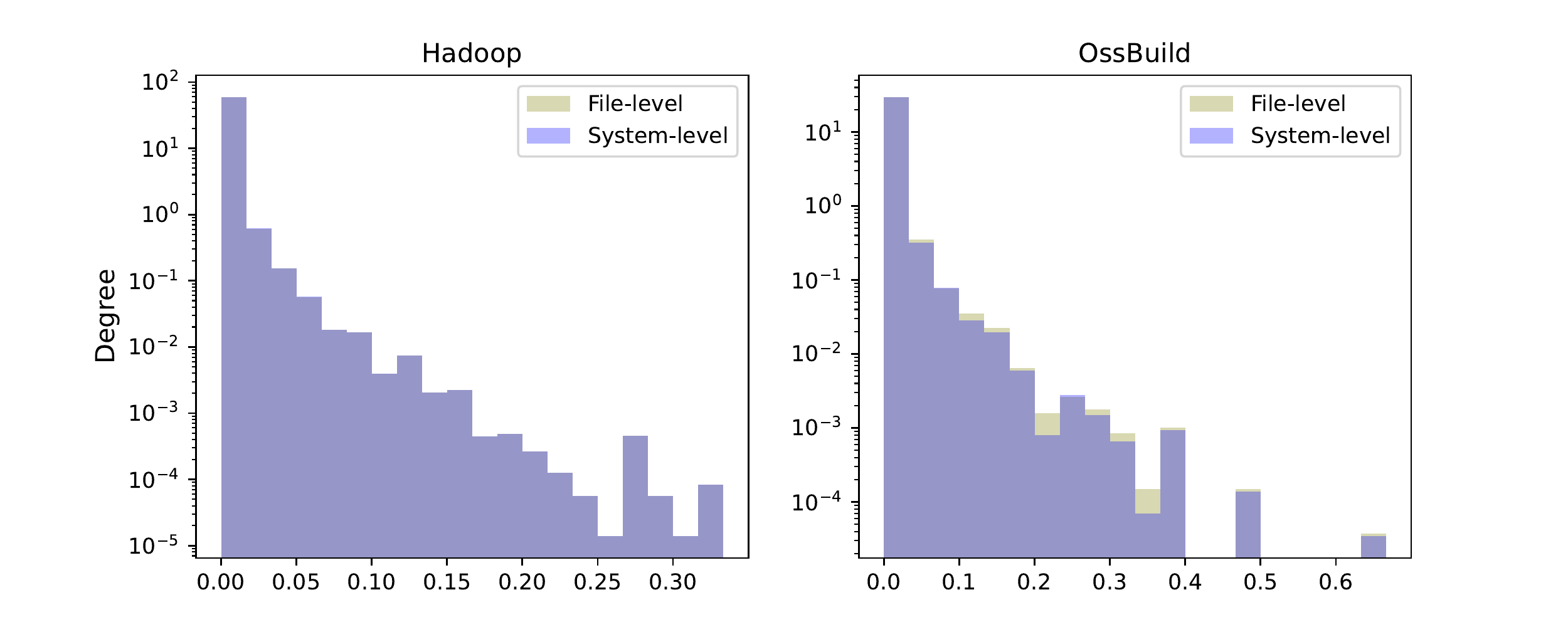}
    \caption{Degree distribution in logarithmic scale of Hadoop (left) and OssBuild (right) for both file-level and system-level. }
    \label{fig:app:deg}
\end{figure}

The degree distribution, depicted in Figure \ref{fig:app:deg}, effectively captures the resemblance between the distributions of nodes and edges.

\subsection{Triangles}

In network science, the concept of triangle closure, also known as the "friendship paradox", is a well-established and widely recognized phenomenon. It has garnered significant attention and has been extensively studied in various research works, highlighting its relevance and importance in numerous real-world applications.
In particular, we explore the relationship between the graph and triangles through Transitivity\cite{watts1998collective} and Clustering Coefficient\cite{newman1999renormalization}.
Transitivity is defined as follows:
\begin{equation}
    \text{Transitivity} = 3\cdot \dfrac{\text{$\#$ of triangles}}{\text{$\#$ of triads}}
\end{equation}
On the other hand, the clustering coefficient is a metric associated with a given node $u$, and it refers to the degree to which nodes in a graph tend to cluster together. The clustering coefficient of a node $u$ is defined as follows:
\begin{equation}
    C_u = \dfrac{2\cdot T(u)}{(deg(u))\cdot(deg(u)-1)}
\end{equation}
Where $T(u)$ is the number of triangles through node $u$, and $deg(u)$ is the degree of node $u$.
The Global Clustering Coefficient (GCC) is the average among the clustering coefficient of all nodes.
In summary, while both transitivity and clustering coefficient capture the local clustering patterns in a network, transitivity focuses on the presence of triangles and overall network connectivity, whereas the clustering coefficient specifically measures the density of connections between neighboring nodes.

\begin{figure}[h!] 
\centering
\subfigure[]{\label{fig:app:tran}\includegraphics[width=\linewidth]{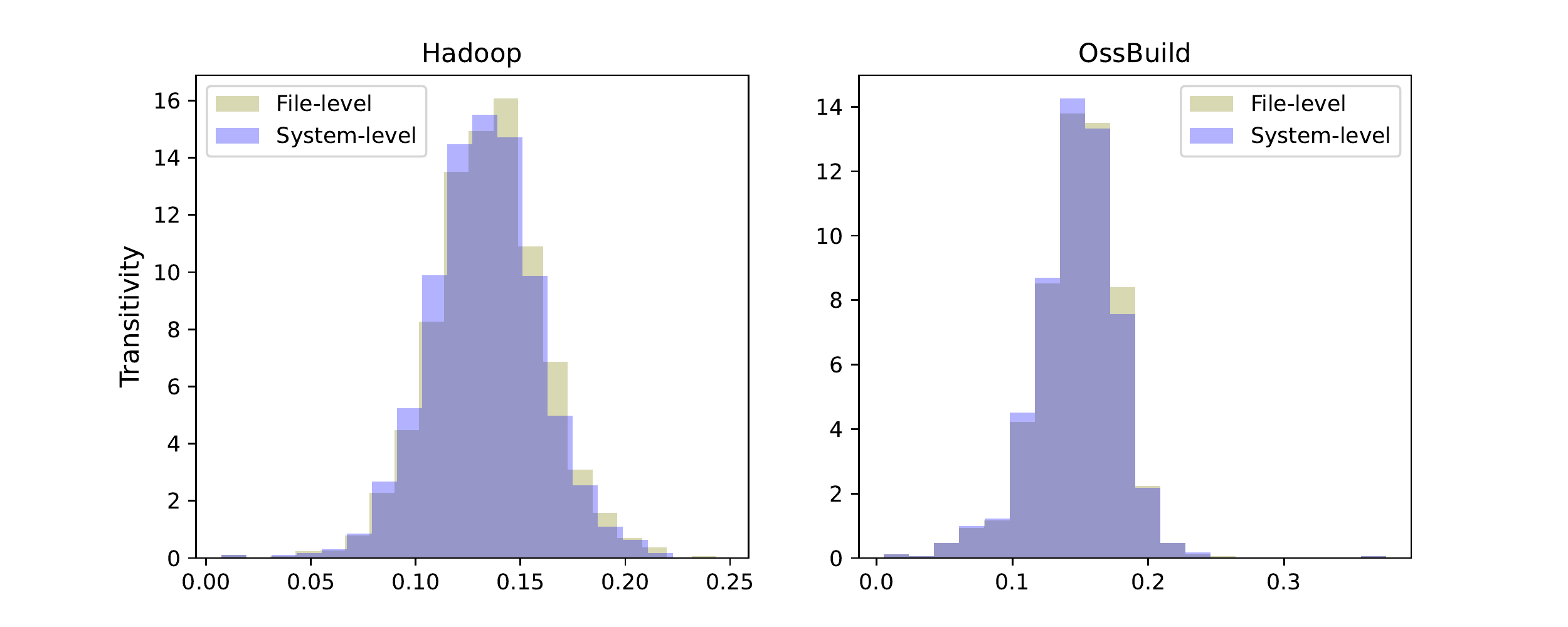}}
\subfigure[]{\label{fig:app:clu}\includegraphics[width=\linewidth]{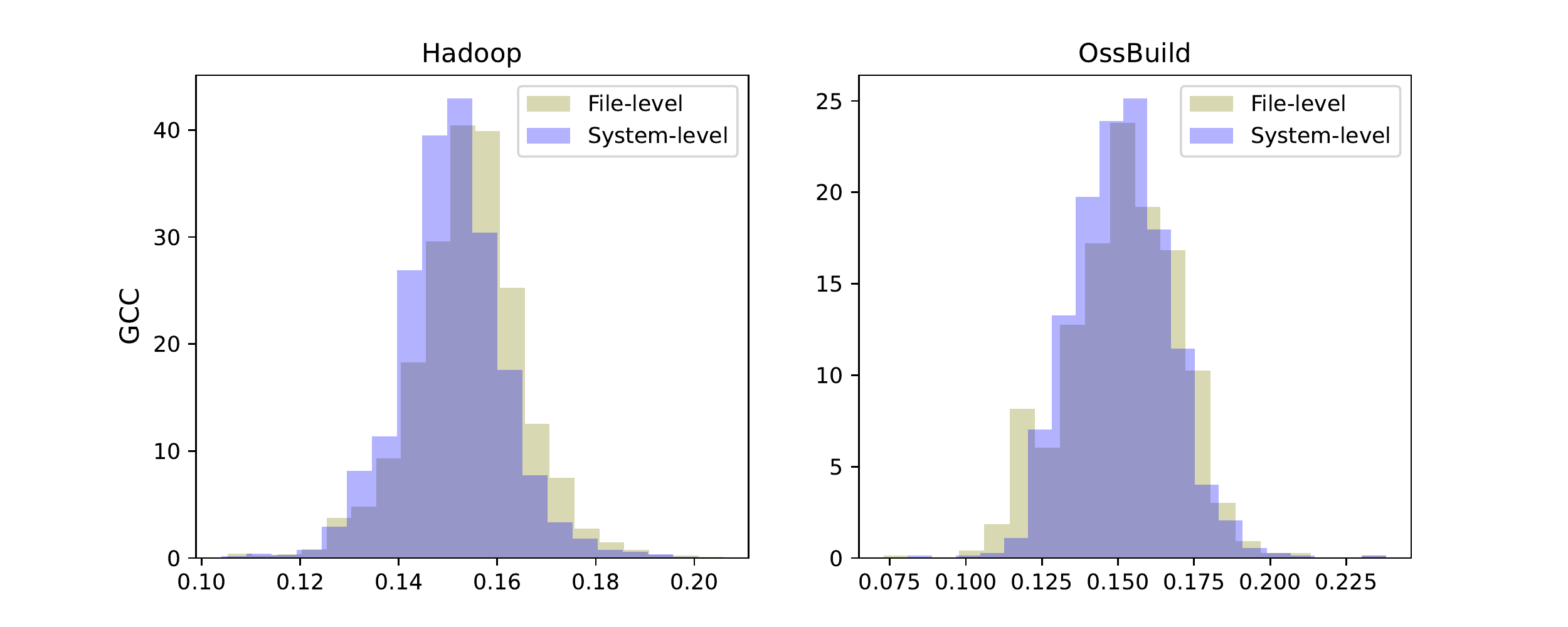}}
\caption{Distributions of the transitivity and global clustering coefficient (GCC) in Hadoop (left) and OssBuild (right) for both file-level and system-level settings.}
\label{fig:app:tranclu}
\end{figure}

Figure \ref{fig:app:tranclu} shows the transitivity (Figure \ref{fig:app:tran}) and the global clustering coefficient (Figure \ref{fig:app:clu}) distributions.
Based on the results, it is apparent that both transitivity and GCC exhibit higher values in the file-level dataset compared to the system-level dataset. However, this distinction is not as pronounced in the OssBuild dataset.

\subsection{Assortativity}

\begin{figure}[h!]
    \centering
    \includegraphics[width=\textwidth]{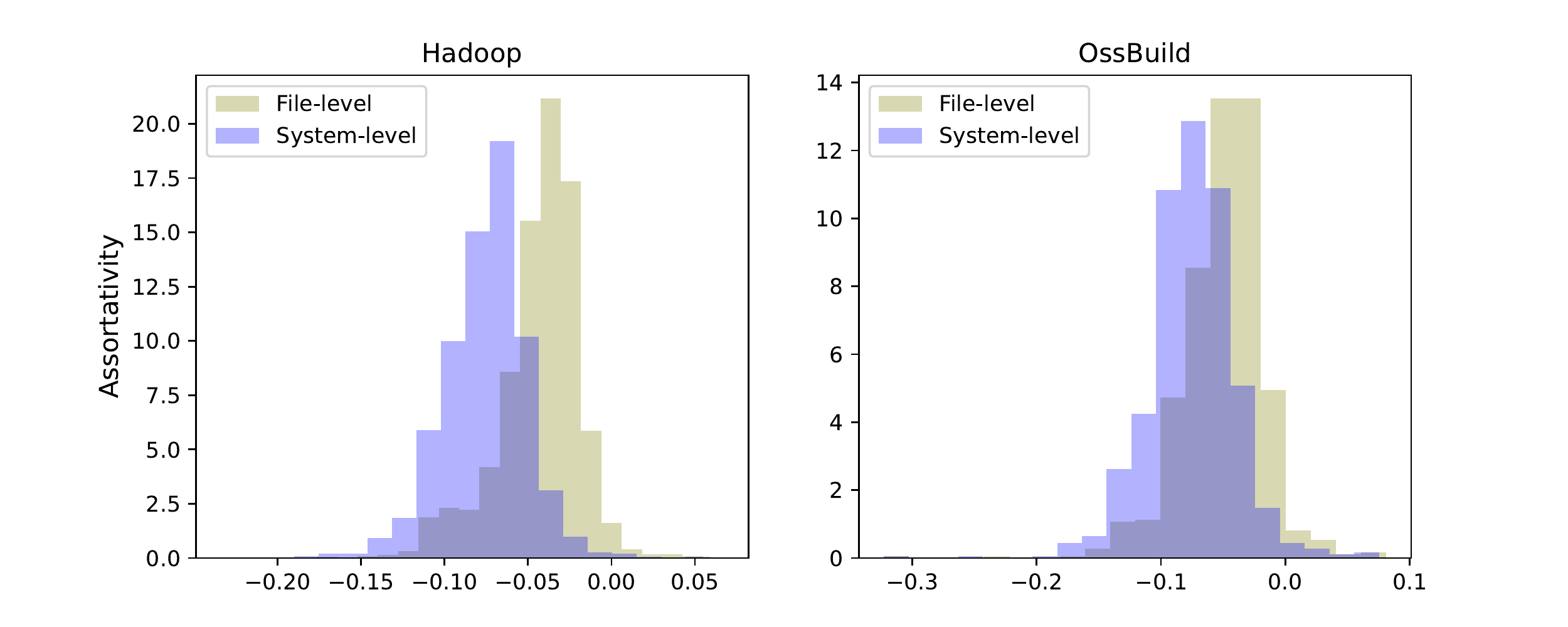}
    \caption{Distributions of the degree assortativity in Hadoop (left) and OssBuild (right) for both file-level and system-level.}
    \label{fig:app:ass}
\end{figure}

Assortativity, in network theory, refers to the tendency of nodes in a network to connect with similar nodes. It measures the degree of homophily or assortative mixing in a network based on node attributes or characteristics. Assortativity can be quantified using various metrics, such as degree assortativity, attribute assortativity, or assortativity coefficient\cite{newman2002assortative}.
In Figure \ref{fig:app:ass} we report the degree assortativity that examines the correlation of node degrees between connected nodes. 

Based on the observations in Figure \ref{fig:app:ass}, it is challenging to determine whether the graphs exhibit positive assortativity (where nodes with similar degrees tend to connect) or negative assortativity (indicating connections between nodes with differing degrees). However, upon examining the histograms, it appears that in both scenarios, the System-level dataset tends to connect nodes to other nodes with differing degrees.

\subsection{Centralities}

Centrality in network analysis refers to the importance or prominence of nodes within a network. It measures the extent to which a node is influential, well-connected, or positioned strategically within the network structure. Centrality measures help identify key nodes that play crucial roles in information flow, influence propagation, and network dynamics.

Various centrality measures exist, where we have already evaluated the degree distributions (in Figure \ref{fig:app:deg}). Here we dig deeper into Betweenness Centrality, Closeness Centrality, and Page Rank.
The Betweenness Centrality measures the control a node has over the flow of information in the network. Formally, it is defined as\cite{freeman1977set}
\begin{equation}
    \text{Betweenness Centrality}_u = \sum_{s,t\in V} \dfrac{\sigma(s,t|v)}{\sigma(s,t)}
\end{equation}
where, $\sigma(s,t)$ is the number of shortest paths between node $s$ and node $t$, while $\sigma(s,t|u)$ is the number of shortest paths between node $s$ and node $t$ passing through node $u$.\\

Closeness Centrality measures the proximity of a node to all other nodes in the network. Formally, it is defined as\cite{wasserman1994social}
\begin{equation}
    \text{Closeness Centrality}_u = \dfrac{n-1}{\sum_{v=1}^{n-1} d(v,u)}
\end{equation}
Where here $n$ is the number of nodes, and $d(v,u)$ is the shortest-path length between node $v$ and node $u$.

Finally, the Page Rank\cite{ma2008bringing} assigns importance to nodes based on the number and quality of incoming links. Nodes with higher Page Rank are considered more influential.\\

\begin{figure}[t!] 
\centering
\subfigure[]{\label{fig:app:bet}\includegraphics[width=0.8\linewidth]{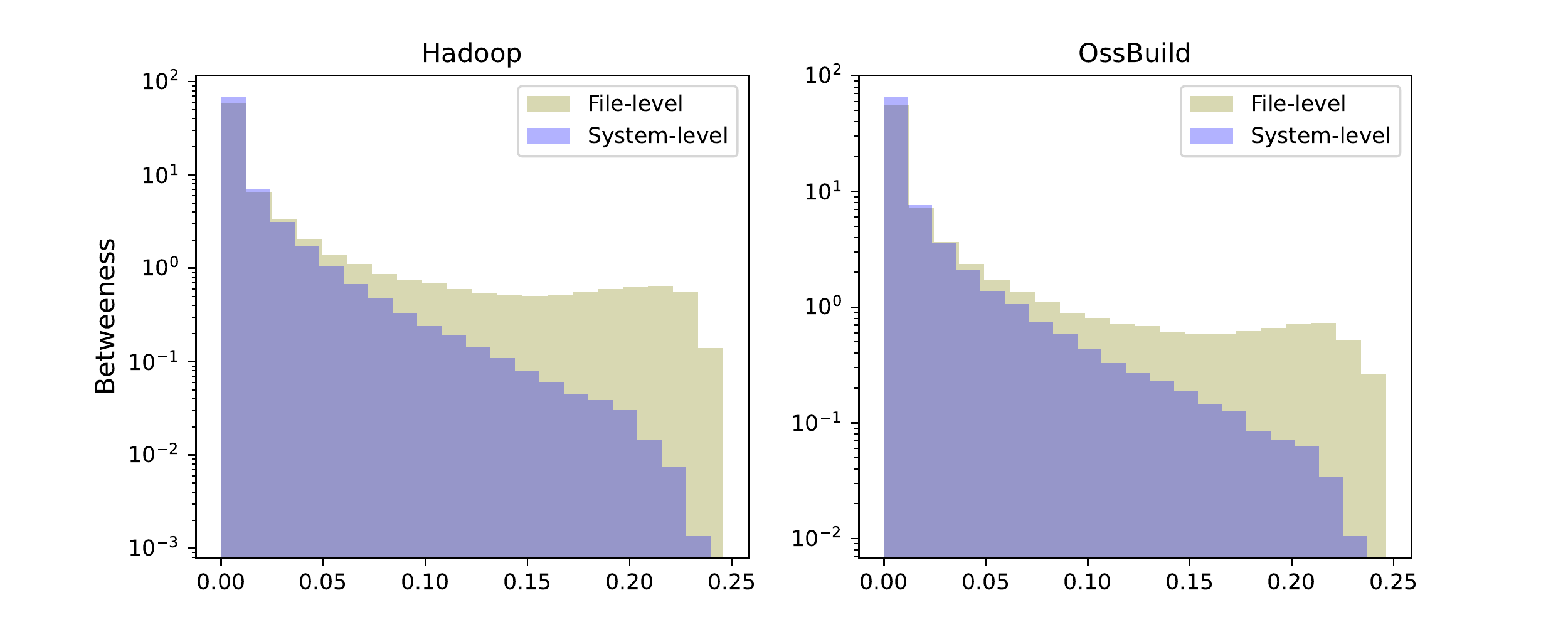}}
\subfigure[]{\label{fig:app:clo}\includegraphics[width=0.8\linewidth]{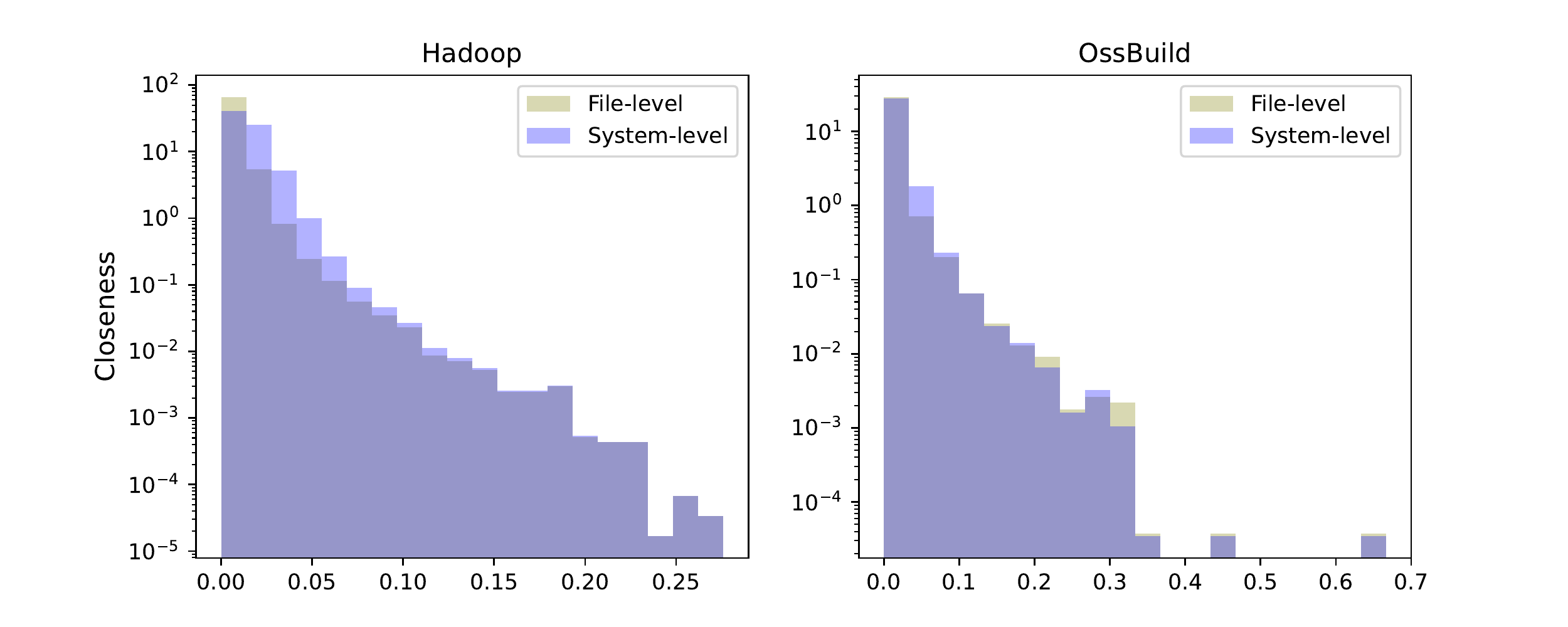}}
\subfigure[]{\label{fig:app:pagrank}\includegraphics[width=0.8\linewidth]{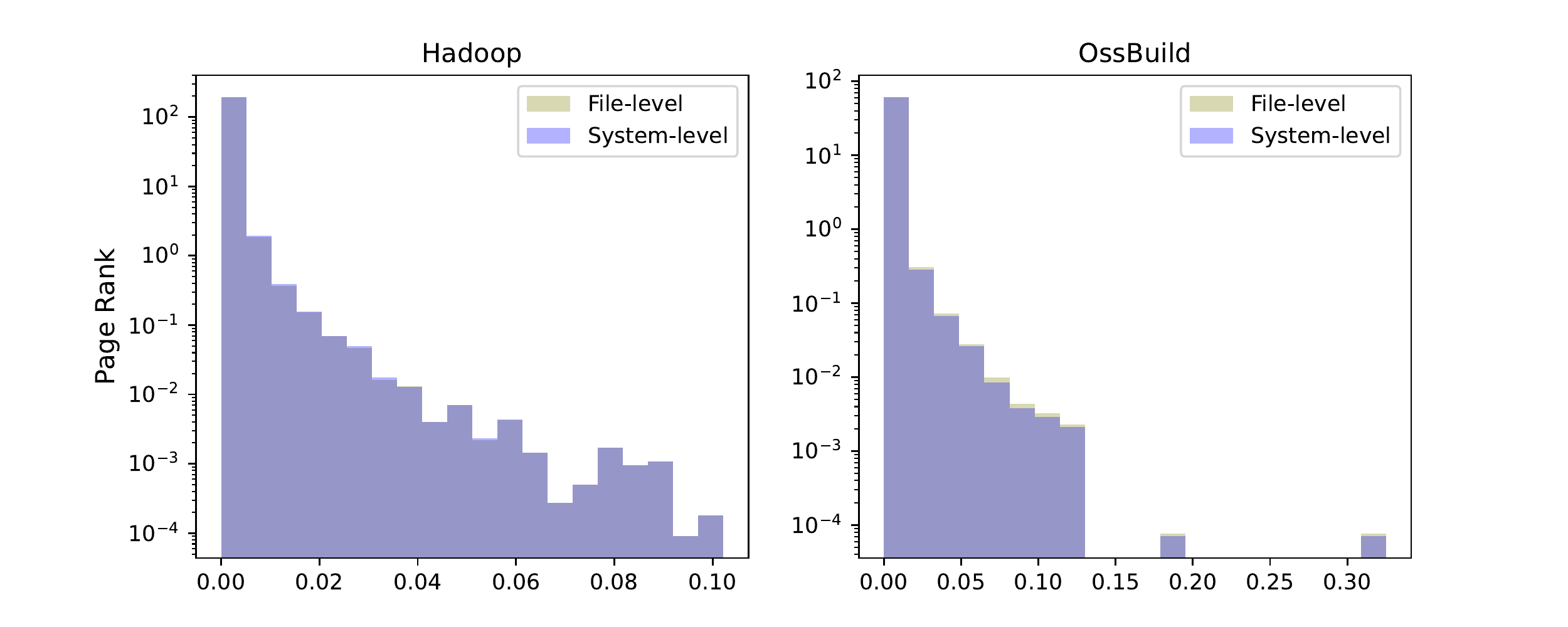}}
\caption{Distributions of the Betweenness, Closeness and Page Rank (in log scale) in Hadoop (left) and OssBuild (right) for both file-level and system-level settings.}
\label{fig:app:betclopag}
\end{figure}

In Figure \ref{fig:app:betclopag} we report the Betweenness, Closeness and Page Rank of the datasets. It is clear that the strongest difference between the file and system level settings relies on the Betweenness. This is not surprising at all, since in the file-level setting there are fewer edges, thus the number of edges with a higher Betweenness is greater.

\subsection{Meso-scale}

In conclusion, we explore the meso-scale characteristics of the network topology by employing measures such as shortest-path analysis\cite{newman2018networks}, tree similarity, and the diameter\cite{newman2018networks} of the input network.

\begin{figure}[h!]
    \centering
    \includegraphics[width=\textwidth]{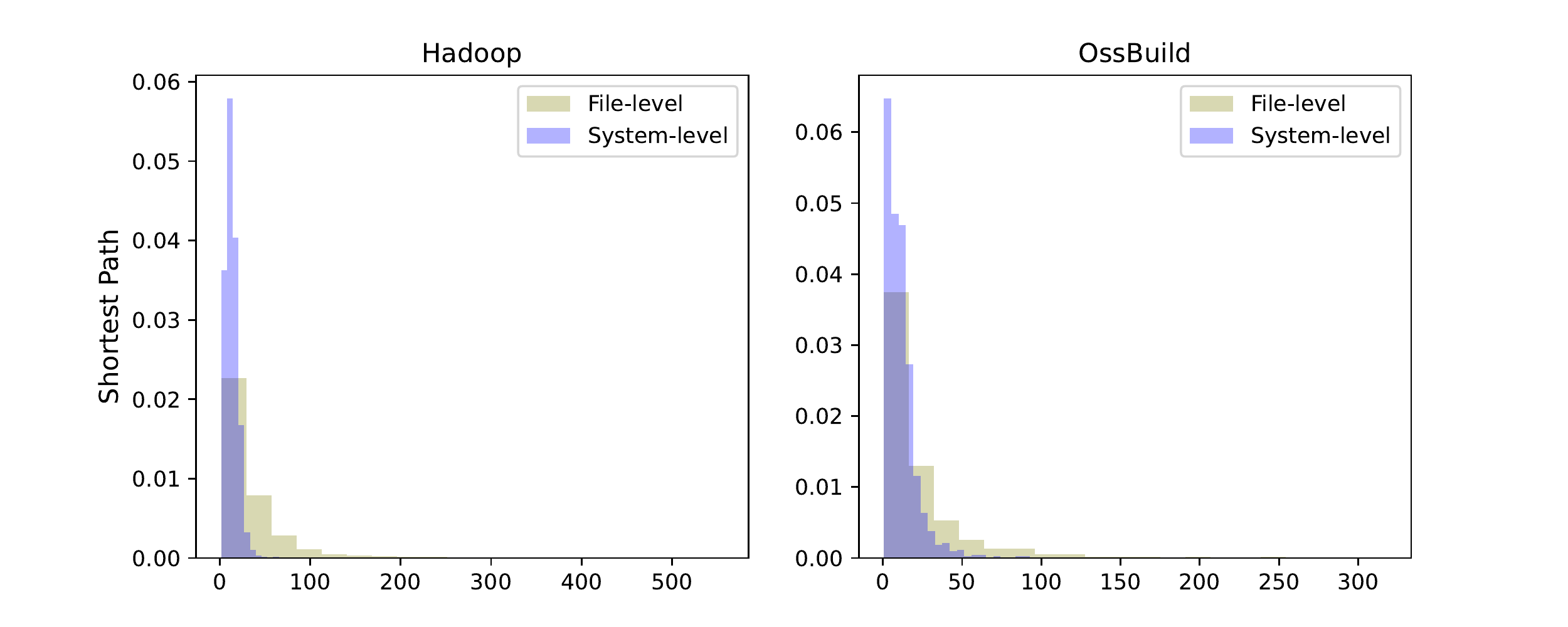}
    \caption{Shortest path length distributions.}
    \label{fig:app:short}
\end{figure}

The shortest path is defined in Definition 2, and it reports the smaller path connection between two given nodes. The distribution is reported in Figure \ref{fig:app:short}.
The figure clearly indicates that the system-level network exhibits shorter shortest paths compared to the file-level network. This observation is expected, as the system-level networks contain a higher number of edges in comparison to the file-level networks.

\begin{figure}[h!]
    \centering
    \includegraphics[width=\textwidth]{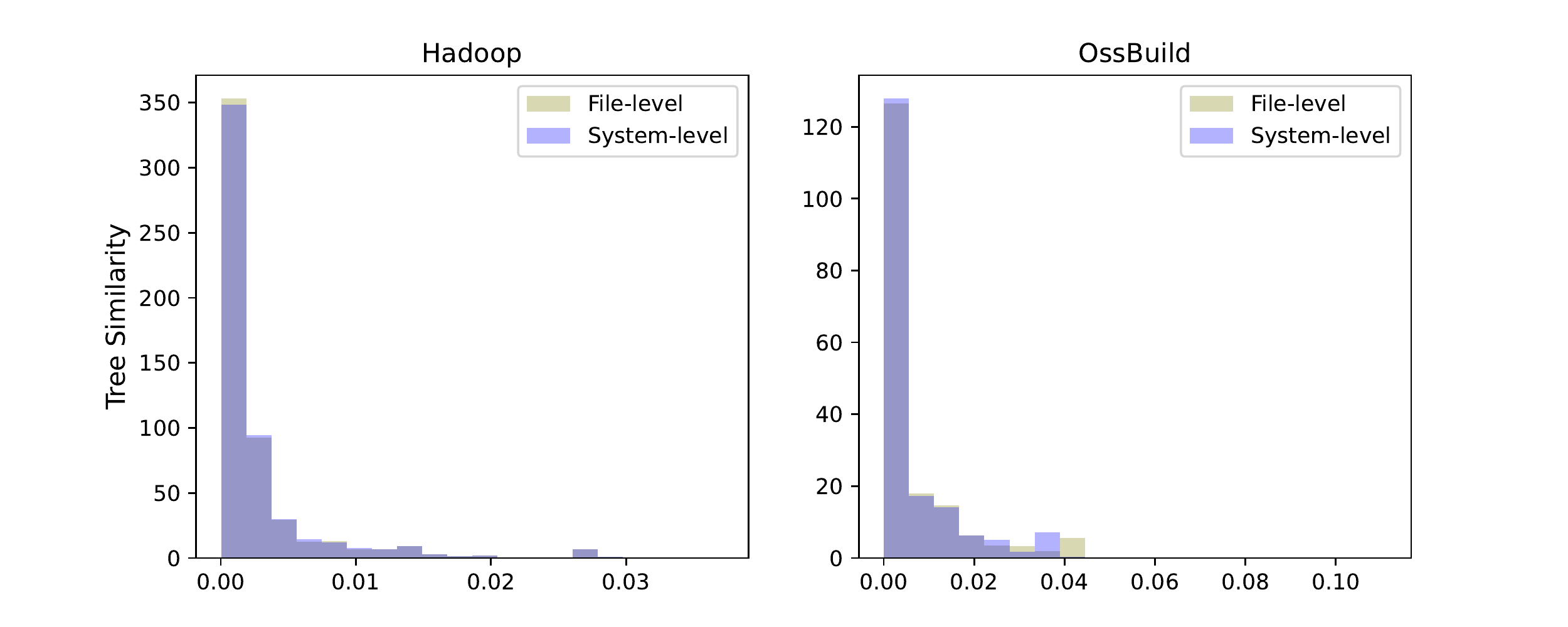}
    \caption{\textit{Tree sim} distributions.}
    \label{fig:app:treesim}
\end{figure}

The \textit{tree-sim} metric, defined in Eq. \ref{eq:treesim}, is a custom measure that quantifies the similarity between the input graph and its corresponding tree structure. It is important to note that this metric should not be confused with Treewidth. In our study, we introduced the tree-sim metric as an alternative to overcome the computational complexity associated with calculating Treewidth. The distribution of the tree-sim metric for each dataset is presented in Figure \ref{fig:app:treesim}. However, no significant insights or noteworthy patterns were observed from the analysis of these distributions, where, as expected, both follow power-law distribution.

\begin{figure}[h!]
    \centering
    \includegraphics[width=\textwidth]{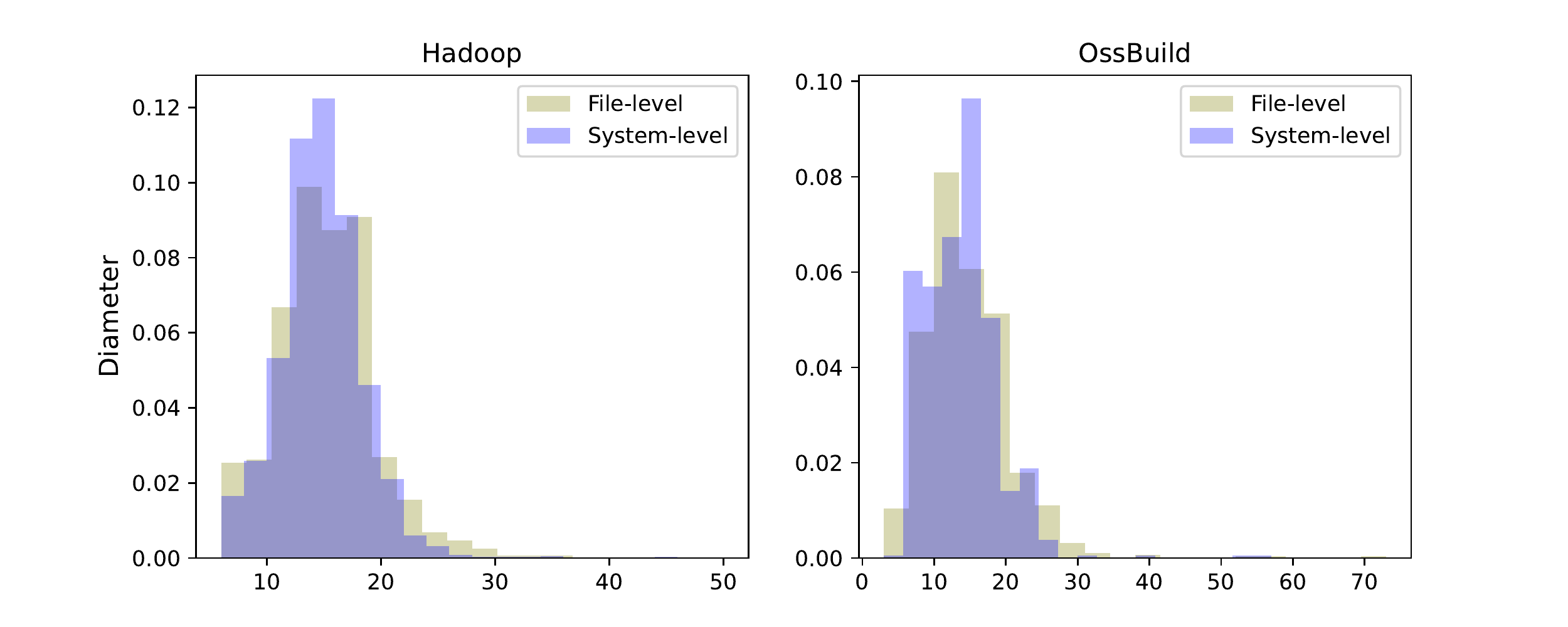}
    \caption{Diameter distributions.}
    \label{fig:app:diam}
\end{figure}

Lastly, in Figure \ref{fig:app:diam}, we present the distribution of diameters for each graph. As expected, the system-level networks exhibit a smaller diameter compared to the file-level networks. 

\newpage

\section{Shallow Embedding Results When Test Features are Not Manipulated}
\label{append_results}
As we mentioned in Section~\ref{sec:passivelearning}, computing the embedding using only the training features without manipulating the test features in embedding is not possible for unsupervised shallow embedding because eventually, we will have different features space for both training and testing data. In this section, we will prove the aforementioned statement for both passive and active learning. 

\subsection{Passive Learning}
We are experimenting with computing the embedding only for one seed for the dataset. We compute the embedding for the entire dataset ( when test features are included) and then we get the first 80\% of the dataset and compute the embedding only for this portion of the dataset( so here, the last 20\% are excluded). Here we either retrain the model based on the last 20\%, which leads to poor results or alternatively, we get the benefit of the embedding of the entire dataset since the training data is included. Then using the same split index that we did when we got the training data, we get the last 20\% of the embedding. 

\subsubsection{System-Level Parsing}

\begin{table}[h!]
\caption{Results for Unsupervised Embedding for graphs of System Level Parsing.}
\label{tab:resultsunsup_appendix}
\centering
\begin{tabular}{@{}c|cc|cc|cc@{}}
 & & & \multicolumn{2}{c|}{Train and Test Features} & \multicolumn{2}{c}{Train Features} \\
 \hline
 & & & OSSBuilds & HadoopTests & OSSBuilds & HadoopTests \\
\hline
 \multirow{9}{*}{\STAB{\rotatebox[origin=c]{-90}{Shallow Embedding}}} & \multicolumn{2}{c|}{Graph2Vec} & \textbf{0.78} & \textbf{0.74} & \textbf{0.65} & {0.46} \\
& \multirow{2}{*}{GR} & mean & 0.44 & 0.49 & 0.50 & 0.45 \\
& & sum & 0.45 & 0.42 & 0.45 &\textbf{ 0.48} \\
& \multirow{2}{*}{HOPE} & mean & 0.14 & 0.015 & 0.16 & 0.04 \\
& & sum & 0.13 & 0.36 & 0.16 & 0.39 \\
& \multirow{2}{*}{DeepWalks} & mean & 0.41 & 0.47 & 0.38 & 0.4 \\
& & sum & 0.43 & 0.45 & 0.32 & 0.39 \\
& \multirow{2}{*}{Node2Vec} & mean & 0.39 & 0.42 & 0.29 & 0.32 \\
& & sum & 0.44 & 0.42 & 0.33 & 0.43 \\
\hline

\end{tabular}
\end{table}

Table~\ref{tab:resultsunsup_appendix}, shows the results for shallow embedding with and without test features with one split for the data. These results are significantly better than the ones we averaged for 15 different splits when we used the test features. However, with more splits, the results are more reliable. Looking at Table~\ref{tab:resultsunsup_appendix}, using only the train data features leads to a substantial decline in the performance of Graph2Vec, DeepWalk, and Node2Vec (except for sum aggregation in the HadoopTest dataset). These methods are all shallow embedding techniques based on skip-gram, as shown in Figure~\ref{fig:unsupervised}. Conversely, there are generally slight improvements for the other shallow methods based on Matrix Factorization, such as HOPE and GR (except for mean aggregation in the HadoopTests dataset). 

\subsubsection{File-Level Parsing}
The results for this setting are reported in Table~\ref{tab:resultsunsup_NOSUT_appendix}. In this setting, we still have better results than those obtained with 15 different splits for the dataset. 
\begin{table}[h!]
\caption{Results for Unsupervised Embedding for graphs of File Level Parsing.}
\label{tab:resultsunsup_NOSUT_appendix}
\centering
\begin{tabular}{@{}c|cc|cc|cc@{}}
 & & & \multicolumn{2}{c|}{Train and Test Features} & \multicolumn{2}{c}{Train Features} \\
 \hline
 & & & OSSBuilds & HadoopTests & OSSBuilds & HadoopTests \\
\hline
 \multirow{9}{*}{\STAB{\rotatebox[origin=c]{-90}{Shallow Embedding}}} & \multicolumn{2}{c|}{Graph2Vec} & \textbf{0.78} & \textbf{0.74} & 0.53 & \textbf{0.49} \\
& \multirow{2}{*}{GR} & mean & 0.57 & 0.46 & \textbf{0.59} & 0.41 \\
& & sum & 0.51 & 0.42 & 0.49 & 0.37 \\
& \multirow{2}{*}{HOPE} & mean & 0.17 & 0.034 & 0.06 & 0.07 \\
& & sum & 0.15 & 0.35 & 0.07 & 0.3 \\
& \multirow{2}{*}{DeepWalks} & mean & 0.45 & 0.43 & 0.34 & 0.24 \\
& & sum & 0.42 & 0.41 & 0.39 & 0.02 \\
& \multirow{2}{*}{Node2Vec} & mean & 0.33 & 0.2 & 0.39 & 0.15 \\
& & sum & 0.33 & 0.36 & 0.31 & 0.32 \\
\hline

\end{tabular}
\end{table}

Nevertheless, when we exclude the test features, the correlation score for Graph2Vec is drastically reduced to 0.53 for OssBuilds and 0.49 for HadoopTests which remains the best for such dataset when we only use the train features. Conversely, GR with mean aggregation is the best for the same setting for OssBuilds. 


\subsection{Active Learning}
To understand the impact of different feature spaces embedding we will present the active learning results for Graph2Vec when test features are not included. 

\subsubsection{System-Level Parsing}

\begin{figure}[htb!] 
\centering
\subfigure[]{\label{subfig:g2v_sys_oss}\includegraphics[width=0.24\linewidth]{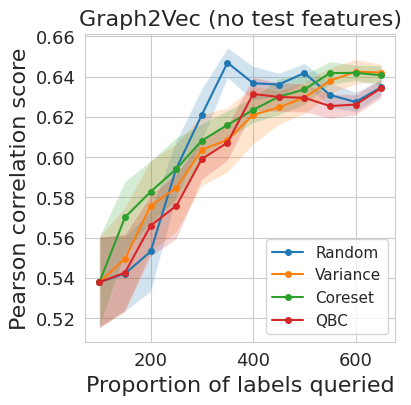}}
\subfigure[]{\label{subfig:g2v_sys_hadoop}\includegraphics[width=0.24\linewidth]{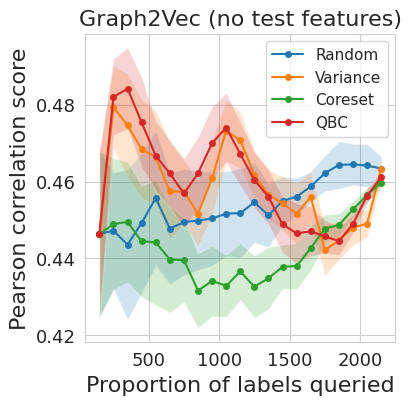}}
\caption{Active learning results for Graph2Vec When Test Features are not Used in embeddings for the OSSBuilds (left) and HadoopTest (right) datasets (System Level Parsing) with $|\mathcal{L}_0| = 100$ and $|\mathcal{B}| = 50$.}
\label{fig:g2v_false_embeddings}
\end{figure}

In Figure~\ref{fig:g2v_false_embeddings}, the embedding performance based on Graph2Vec without test features for HadopTests only improves slightly at the start but then stays fairly constant. The reason for this is likely because the resulting latent graph representation is not rich enough for this embedding past 500 labels. We have the same issue for the OSSBuilds dataset for random and QBC. 

\subsubsection{File-Level Parsing}
\begin{figure}[htb!] 
\centering

\subfigure[]{\label{subfig:g2v_file_oss}\includegraphics[width=0.24\linewidth]{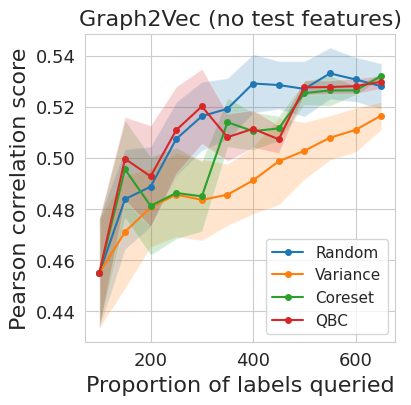}}
\subfigure[]{\label{subfig:g2v_file_hadoop}\includegraphics[width=0.24\linewidth]{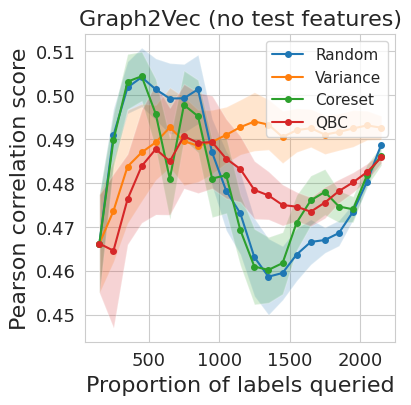}}
\caption{Active learning results for Graph2Vec When Test Features are not Used in embeddings for the OSSBuilds (left) and HadoopTest (right) datasets (File Level Parsing) with $|\mathcal{L}_0| = 100$ and $|\mathcal{B}| = 50$.}
\label{fig:g2v_false_embeddings_NOSUT}
\end{figure}

In Graph2Vec with no test features in Figure~\ref{fig:g2v_false_embeddings_NOSUT}, for the HadoopTests dataset, coreset and random are the best choice when we have up to 1000 samples but the quality of labelling drastically reduces after that threshold when variance remains the best as it performs reliably after 500 samples. Variance is the worst option for the OSSBuilds dataset. 

\end{document}